\documentstyle[12pt]{article} \voffset=0mm \headheight=0mm
\date{}
\author{Valerii Dryuma\thanks{Work supported in part by MURST, Italy}\\[5mm]
{\it Institute of Mathematics and Informatics, AS RM,}\\[3mm]
{\it 5 Academiei Street, 2028 Kishinev, Moldova},\\[3mm]{\it e-mail:
valery@gala.moldova.su} }
\title{THE RIEMANN EXTENSIONS IN THEORY OF ORDINARY DIFFERENTIAL EQUATIONS
AND THEIR APPLICATIONS}

\newtheorem{pr}{Proposition}
\newtheorem{rem}{Remark}
\begin{document}
\maketitle
\date{}
\maketitle
\begin{abstract}
Some properties of the 4-dim Riemannian spaces with the metrics
$$
ds^2=2(za_3-ta_4)dx^2+4(za_2-ta_3)dxdy+2(za_1-ta_2)dy^2+2dxdz+2dydt
$$
 associated with the second order nonlinear differential equations
$$
y''+a_{1}(x,y){y'}^3+3a_{2}(x,y){y'}^2+3a_{3}(x,y)y'+a_{4}(x,y)=0
$$
with arbitrary coefficients $a_{i}(x,y)$ and 3-dim Einstein-Weyl
spaces connected with dual equations
$$
b''=g(a,b,b')
$$
where the function $g(a,b,b')$ is satisfied the partial differential equation
$$
g_{aacc}+2cg_{abcc}+2gg_{accc}+c^2g_{bbcc}+2cgg_{bccc}+
g^2g_{cccc}+(g_a+cg_b)g_{ccc}-4g_{abc}-
$$
$$
 -4cg_{bbc} -cg_{c}g_{bcc}-
3gg_{bcc}-g_cg_{acc}+ 4g_cg_{bc}-3g_bg_{cc}+6g_{bb} =0
$$
are considered.

The theory of the invariants of second order ODE's for
investigation of the nonlinear dynamical systems with parameters
is used. The applications to the Riemann spaces in General
Relativity are discussed.
\end{abstract}

\section
{Introduction}

The  second order ODE's of the type
$$
y''+a_1(x,y){y'}^3+3a_2(x,y){y'}^2+3a_3(x,y)y'+
a_4(x,y)=0  \eqno (1)
$$
are connected with  nonlinear dynamical systems in the form
$$
\frac{dx}{dt}= P(x,y,z,\alpha_i),\quad \frac{dy}{dt}= Q(x,y,z,\alpha_i),
\quad \frac{dz}{dt}= R(x,y,z,\alpha_i),
$$
where $\alpha_i$ are parameters.

For example the Lorenz system
$$
\dot X=\sigma(Y-X),\quad \dot Y=rX-Y-ZX,\quad \dot Z=XY-bZ
$$
having  chaotic properties at some values of parameters
is equivalent to the equation
$$
y''-\frac{3}{y}{y'}^2+(\alpha y - \frac{1}{x})y'+\epsilon xy^4-\beta x^3 y^4-
\beta x^2y^3-\gamma y^3 + \delta \frac{y^2}{x}=0, \eqno (2)
$$
where
$$
\alpha=\frac{b+\sigma+1}{\sigma},\quad \beta=\frac{1}{\sigma^2},
\quad \gamma=\frac{b(\sigma+1)}{\sigma^2},\quad \delta=\frac{(\sigma+1)}{\sigma},
\quad \epsilon=\frac{b(r-1)}{\sigma^2},
$$
and for investigation of its properties  the theory of invariants
was first used in [1--5].

     Other example is the  third order differential equation
$$
\frac{d^3 X}{d t^3}+a \frac{d^2 X}{d t^2}-\left(\frac{d X}{d t}\right)^{2}+X=0   \eqno(3)
$$
with parameter $a$ having chaotic properties at the values $2,017<
a <2,082$ [25]. It can be transformed to the form (1)
$$
y''+\frac{1}{y}y'^2+\frac{a}{y}y'+\frac{x}{y^2}-1=0
$$
with the help of standard substitution
$$
\frac{d X}{d t}=y(x),\quad \frac{d^2 X}{d t^2}=y'y,\quad
\frac{d^3 X}{d t^3}=y''y^2+y'^2 y.
$$

   According to the Liouville theory [6--10] all equations of
type (1) can be devided in  two different classes

I. \ $\nu_5=0$,

II. \ $\nu_5\neq0$.

   Here the value $\nu_5$ is the expression of the form
$$
\nu_5=L_2(L_1L_{2x}-L_2L_{1x})+L_1(L_2L_{1y}-L_1L_{2y})-a_1L_1^3+3a_2L_1^
2L_2-3a_3L_1L_2^2+a_4L_2^3\,
$$
then $ L_1,L_2 $ are defined by formulaes
$$
L_1=\frac{\partial}{\partial y}(a_{4y}+3a_2a_4)-\frac{\partial}{\partial x}
(2a_{3y}-a_{2x}+a_1 a_4)-3a_3(2a_{3y}-a_{2x})-a_4a_{1x} ,
$$
$$
L_2=\frac{\partial}{\partial x}(a_{1x}-3a_1a_3)+\frac{\partial}{\partial y}
(a_{3y}-2a_{2x}+a_1a_4)-3a_2(a_{3y}-2a_{2x})+a_1a_{4y} .
$$

   For the equations with condition $\nu_5 =0$
R. Liouville  discovered  the series of semi-invariants
starting from :
$$
w_1={1\over L_1^4}\left[L_1^3(\alpha' L_1-\alpha'' L_2)+R_1(
L_1^2)_x-L_1^2R_{1x}+L_1R_1(a_3L_1-a_4L_2)\right] ,
$$
where
$$
 R_1= L_1L_{2x}-L_2L_{1x}+a_2 L_1^2-2 a_3 L_1L_2 +a_4 L_2^2
$$
or
$$
w_2={1\over L_2^4}\left[L_2^3(\alpha' L_2-\alpha L_1)-R_2(
L_2^2)_y+L_2^2R_{2y}-L_2R_2(a_1L_1-a_2L_2)\right] ,
$$
where
$$
 R_2= L_1L_{2y}-L_2L_{1y}+a_1 L_1^2-2 a_2 L_1L_2 +a_3 L_2^2
$$
and
$$
 \alpha=a_{2y}-a_{1x}+2(a_1 a_3-a_2^2), \quad
 \alpha'=a_{3y}-a_{2x}+a_1 a_4-a_2a_3,
$$
$$
 \alpha''= a_{4y}-a_{3x}+2(a_2 a_4-a_3^2).
$$
It has the form
$$
w_{m+2}=L_1\frac{\partial w_m}{\partial y} -L_2\frac{\partial w_m}
{\partial x}+mw_m(\frac{\partial L_2}{\partial x}-\frac{\partial L_1}
{\partial y}).
$$
  In case $w_1=0$ there are another series of semi-invariants
$$
i_{2m+2}=L_1\frac{\partial i_{2m}}{\partial y} -L_2\frac{\partial i_{2m}}
{\partial x}+2mi_{2m}(\frac{\partial L_2}{\partial x}-\frac{\partial L_1}
{\partial y}).
$$
where
$$
i_2={3R_1\over L_1}+{\partial L_2\over\partial x}-{\partial L_1\over
\partial y} .
\label{i2}
$$
and corresponding sequence for absolute invariants
$$
j_{2m}=\frac{i_{2m}}{i_2^{m}}.
$$

    In case $\nu_5\neq0$  the semi-invariants have the form
$$
\nu_{m+5}=L_1\frac{\partial \nu_{m}}{\partial y} -L_2\frac{\partial \nu_{m}}
{\partial x}+m\nu_{m}(\frac{\partial L_2}{\partial x}-\frac{\partial L_1}
{\partial y}).
$$
and corresponding series of absolute invariants
$$
[5t_{m}-(m-2)t_7t_{m-2}]\nu_5^{2/5}=5(L_1\frac{\partial t_{m-2}}{\partial y} -
L_2\frac{\partial t_{m-2}}{\partial x}) \eqno(4)
$$
where
$$
t_m=\nu_m \nu_5^{-m/5}
$$

 From the formulaes (4) follows that some relations between
the invariants are important for theory of second order ODE.

In fact, let
$$
t_9=f(t_7)
$$
be the example of such relation.
Then we have
$$
(5f(t_7)-7t_7^2)\nu_5^{2/5}=5(L_1\frac{\partial t_7}{\partial y}-L_2\frac{\partial t_7}{\partial x}),
$$
$$
(5(t_{11}-9t_7f(t_7))\nu_5^{2/5}=5f'(t_7)(L_1\frac{\partial t_7}{\partial y}-
L_2\frac{\partial t_7}{\partial x}),
$$
$$
(5t_{13}-11t_7t_{11})\nu_5^{2/5}=5(L_1\frac{\partial t_{11}}{\partial y}-
L_2\frac{\partial t_{11}}{\partial x}),
$$
and
$$
(5f-7t_7^2)f'_{t7}=5t_{11}-9f(t_7)t_7,
$$
from which we get  $t_{11}=g(t_7)$.

In the simplest case $t_9=a t_7^2$ we have $$
t_{11}=a(2a-1)t_7^3,\quad t_{13}=a(2a-1)(3a-2)t_7^4,\quad
t_{15}=a(2a-1)(3a-2)(4a-3)t_7^5\quad... $$

These relations show that some value of parameters
$$
a=0,\quad 1/2,\quad 2/3,\quad 3/4,\quad 4/5...
$$
are special for the corresponding second order ODE's.

To take the example of equation in form
$$
y''+a_1(x,y)y'^3+a_2(x,y)y'^2+3(-xa_2(x,y)-ya_1(x,y))y'+
$$
$$
(x^2-y)a_2(x,y)+xya_1(x,y)-2/3=0.
$$

In the case
$$
 a_1(x,y)=-\frac{2x}{5y^2},\quad a_2(x,y)=\frac{2}{5y}
$$
 we get a following expressions for the relative invariants
 $$
  \nu_5=\frac{4194304}{6328125}\frac{x^3}{y^9},\quad
  \nu_7=-\frac{134217728}{158203125}\frac{x^3}{y^{12}},\quad
   \nu_9=\frac{8589934592}{3955078125}\frac{x^3}{y^{15}},
$$
 $$
 \nu_{11}=-\frac{274877906944}{32958984375}\frac{x^3}{y^{18}},
  \quad
  \nu_{13}=\frac{35184372088832}{823974609375}\frac{x^3}{y^{21}}
  ...,
$$
which corresponds to the series of absolute invariants
$$
\frac{t_9}{t_7^2}=2,\quad \frac{t_11}{t_7^3}=6,\quad
\frac{t_{13}}{t_7^4}=24,\quad \frac{t_{15}}{t_{7}^5}=120,\quad
\frac{t_{17}}{t_7^6}=720,\quad \frac{t_{19}}{t_{7}^7}=5040,
$$
$$
 \frac{t_{21}}{t_{7}^8}=40320,\quad
\frac{t_{23}}{t_{7}^9}=362880,\quad
 \frac{t_{25}}{t_7^{10}}=3628800.
$$

So we have the example of the second order ODE with the invariants
forming the series
$$
 2,\quad 6, \quad 24, \quad
120, \quad 720, \quad 5040, \quad 40320, \quad 362880, \quad
3628800, \quad ...
 $$

It is interesting to note that the ratios of neighbouring  members
of such series are the integer numbers
$$
 3,\quad 4, \quad 5,
\quad 6, \quad 7, \quad 8, \quad 9, \quad 10, \quad ....
$$

   Note that the first applications of the Liouville theory for the
studying of the properties of nonlinear dynamical systems like
Lorenz system was done in the works of author [1-5].

     In particular for the second order differential equation (2)
equivalent to the Lorenz system the $\nu_5$-invariant has the form
$$
\nu_5=Ax^2+\frac{B}{x^2y^2}+C
$$
where
$$
A=\alpha\beta(10\alpha-\alpha^2-6\delta),\quad B=\alpha(\frac{4}{9}\alpha^2+
\frac{2}{3}\alpha\delta-2\delta^2),\quad C=\alpha(\frac{2}{9}\alpha^4+
6\epsilon\delta-4\alpha\epsilon-\alpha^2\gamma)
$$
In this case the condition
$$
\nu_5=0
$$
corresponds to the conditions
$$
A=0,\quad B=0,\quad C=0
$$
 which contains for example the values
$$
\sigma=-1/5,\quad b=-16/5,\quad r=-7/5
$$
which have not been previously met in theory of the Lorenz system.

   The consideration of the invariants $\nu_{m+2}$ is connected with
unwieldy calculations and does not give us the possibility to
apply them for investigation of this system.

   Here we show that in case of the equation (3) is possible to get
more detailed information.

    With this aim we transform the equation
$$
y''+\frac{1}{y}y'^2+\frac{a}{y}y'+\frac{x}{y^2}-1=0
$$
to the more convenient form.

   At the first steep we find variable $x$ from this equation
$$
x=y^2y''-yy'^2-ayy'+y^2
$$
and after differentiating it we  get the third order ODE which can
be reduced to the second order ODE
$$
y''+\frac{1}{y}y'^2+(\frac{4}{x}+\frac{4}{xy})y'+\frac{a}{x^2}-\frac{2}{xy}+
\frac{1}{x^2y^2}+\frac{y}{x^2}=0.
$$

 For this equation we get the invariants
$$
L_1=\frac{(3y+2a)}{3x^2y^2},\quad L_2=\frac{a}{xy^3},
$$
and
$$
\nu_5=-\frac{1}{9}a^3\frac{(2a^2y+18xy-9)}{x^5y^4},\quad
\nu_7=\frac{1}{27}a^4\frac{(54xy^2-27y-20a^3y-180axy+72a)}{x^7y^{15}},
$$
$$
\nu_9=\frac{2}{81}a^6\frac{(702xy^2-297y-140a^3y-1260axy+432a)}{x^9y^{19}},
$$
$$
\nu_{11}=\frac{4}{27}a^8\frac{(990xy^2-369y-140a^3y+1260axy+384a)}{x^{11}y^{23}}.
$$
$$
\nu_{13}=\frac{40}{81}a^{10}\frac{(2754xy^2-927y-308a^3y-2772axy+768a)}{x^{13}y^{27}},
$$
$$
\nu_{15}=\frac{80}{243}a^{12}\frac{(42714xy^2-13203y-4004a^3y-36036axy+9216a)}{x^{15}y^{31}},
$$
$$
\nu_{25}=\frac{985600}{6561}a^{22}\frac{(48428550xy^2-11175165y-2704156a^3y-
24337404axy+4718592a)}{x^{25}y^{51}}.
$$

    From these expressions we can see that only numerical values of coefficients
in formulas for invariants are changed at the transition from $\nu_m$ to
$\nu_{m+2}$.

   This fact can be of use for studying the relations between the invariants when
the parameter $a$ is changed. Remark that the starting equation (3) is
connected with the Painleve I equation in case $a=0$.

     Note that the first applications of the Liouville invariants for the
     Painleve equations was done in the works of author [1-6].
      In particular for the
equations of the Painleve type the invariant $\nu_5=0$ and
$w_1=0$.

 As example for the PI equation
 $$ y''=y^2+x
 $$
  presented in the new coordinates $u=y^2-x,\quad v=y $
   $$
v''+(2+8uv^3)v'^3-12uv^2v'^2+6uvv'-u=0
$$
 we get
 $$
L_1=-2,\quad L_2=4v,\quad \nu_5=0, \quad w_1=0.
$$

 Last time
the  relations between the invariants for the all P-type equation
have been studied in the article [30].

\section{The Riemann spaces in theory of ODE's}

Here we present the construction of the Riemann spaces connected
with the equations of type (1).

    We start from the equations of geodisical lines
of two-dimensional space $A_2$ equipped with affine (or Riemann)
connection. They have the form
$$
\ddot x +\Gamma^1_{11} \dot x^2 +2 \Gamma^1_{12} \dot x \dot y+
\Gamma^1_{22} \dot y^2 =0,
$$
$$
\ddot y +\Gamma^2_{11} \dot x^2 +2 \Gamma^2_{12} \dot x \dot y +
\Gamma^2_{22} \dot y^2 =0.
$$
 This system of equations is equivalent to the  equation
$$
  y''-\Gamma^1_{22}{y'}^3+(\Gamma^2_{22}-2\Gamma^1_{12}){y'}^2+
(2\Gamma^2_{12}-\Gamma^1_{11})y'+\Gamma^2_{11}=0
$$
of type (1) but with special choice of coefficients  $a_i(x,y)$.

  The following proposition is valid
\begin{pr}
The equation (1) with the coefficients $a_i(x,y)$ the geodesics
on the surface with the metrics is determined
$$
ds^2=\frac{1}{\Delta^2}[\psi_1dx^2+2\psi_2dxdy+\psi_3dy^2],
$$
where $\Delta=\psi_1\psi_3-\psi_2^{2}$, when the relations
$$
\psi_{1x}+2a_3\psi_1-2a_4\psi_2=0,
$$
$$
\psi_{3y}+2a_1\psi_2-2a_2\psi_3=0,
$$
$$
\psi_{1y}+2\psi_{2x}-2a_3\psi_2+4a_2\psi_1-2a_4\psi_3=0,
$$
$$
\psi_{3x}+2\psi_{2y}+2a_2\psi_2-4a_3\psi_3+2a_1\psi_1=0.
$$
between the coefficients $a_i(x,y)$ and the components of metrics
$\psi_i(x,y)$ are fulfilled.
\end{pr}

   The equations (1) with arbitrary coefficients $a_i(x,y)$ may be considered
as equations of geodesics of 2-dimensional space $A_2$
$$
\ddot x -a_3 \dot x^2 -2a_2 \dot x \dot y-a_1 \dot y^2 =0,
$$
$$
\ddot y +a_4 \dot x^2 + 2a_3 \dot x \dot y + a_2 \dot y^2 =0
$$
equipped with the projective connection  with components
$$
\Pi_1=\left |\begin{array}{cc}
-a_3 & -a_2 \\
a_4 & a_3
\end{array} \right |,
\quad
\Pi_2=\left |\begin{array}{cc}
-a_2 & -a_1 \\
a_3 & a_2
\end{array} \right |.
$$

The curvature tensor of this type of connection  is
$$
R_{12}= \frac{\partial \Pi_2}{\partial x}-
\frac{\partial \Pi_1}{\partial y}+\left[\Pi_1,\Pi_2\right]
$$
and has the components
$$
R^{1}_{112}=a_{3y}-a_{2x}+a_1a_4-a_2a_3=\alpha',\quad
R^{1}_{212}=a_{2y}-a_{1x}+2(a_1a_3-a_{2}^2)=\alpha,
$$
$$
R^{2}_{112}=a_{3x}-a_{4y}+2(a_{3}^2-a_2a_4)=-\alpha'',\quad
R^{2}_{212}=a_{2x}-a_{3y}+a_3a_2-a_1a_4=-\alpha'.
$$

For construction of the Riemannian space connected with the equation
of type (1) we use the notice of Riemannian extension  $W^4$ of space $A_2$
with connection $\Pi^k_{ij}$  [12] . The corresponding  metric is
$$
ds^2=-2\Pi^k_{ij}\xi_k dx^i dx^j+2d\xi_idx^i
$$
and in our case it takes the following form ($\xi_1=z, \xi_2=\tau$)
$$
ds^2=2(z a_3-\tau a_4)dx^2+4(z a_2-\tau a_3)dx dy +2(z a_1-\tau a_2)dy^2
+2dx dz +2dy d \tau. \eqno(5)
$$
So, it is possible to formulate the following statement

\begin{pr}

    For a given equation of type (1) there exists the Riemannian space with
metrics (5) having integral curves of such type of equation as part of its
geodesics.
\end{pr}

Really, the  calculation of geodesics of the space $W^4$ with the metric (5)
lead to the system of equations
$$
\frac{d^2 x}{ds^2} -a_3 \left(\frac{d x}{ds}\right)^2-
2 a_2\frac{d x}{ds}\frac{d y}{ds}-a_1\left(\frac{d y}{ds}\right)^2=0,
$$
$$
\frac{d^2 y}{ds^2} +a_4 \left(\frac{d x}{ds}\right)^2+
2 a_3\frac{d x}{ds}\frac{d y}{ds}+a_2\left(\frac{d y}{ds}\right)^2=0,
$$
$$
\frac{d^2 z}{ds^2} +[z(a_{4y}-\alpha'')-
\tau a_{4x}] \left(\frac{dx}{ds}\right)^2+
2[za_{3y}- \tau (a_{3x}+\alpha'')] \frac{dx}{ds}\frac{dy}{ds}+
$$
$$
+[z(a_{2y}+\alpha)- \tau (a_{2x}+2\alpha')]\left(\frac{d y}{ds}\right)^2+
2a_3 \frac{dx}{ds}\frac{dz}{ds}-2a_4 \frac{dx}{ds}\frac{d \tau}{ds}+
2a_2 \frac{dy}{ds}\frac{dz}{ds}-2a_3 \frac{dy}{ds}\frac{d \tau}{ds}=0,
$$
$$
\frac{d^2 \tau}{ds^2}+[z(a_{3y}-2\alpha')- \tau (a_{3x}-\alpha'')]
\left(\frac{dx}{ds}\right)^2+
2[z(a_{2y}-\alpha)-\tau a_{2x}] \frac{dx}{ds}\frac{dy}{ds}+
$$
$$
+[za_{1y}-\tau(a_{1x}+\alpha)] \left(\frac{dy}{ds}\right)^2+
2a_2 \frac{dx}{ds}\frac{dz}{ds}-2a_3 \frac{dx}{ds}\frac{d \tau}{ds}+
2a_1 \frac{dy}{ds}\frac{dz}{ds}-2a_2 \frac{dy}{ds}\frac{d \tau}{ds}=0.
$$
in which the first two equations of the system for coordinates $x$, $y$
are equivalent to the equation (1).

  In turn two last equations of the system  for coordinates
$z(s)$ and $t(s)$ have the form of the $2\times$ matrix linear second order
differential equations
$$
\frac{d^2 \Psi}{ds^2}+A(x,y)\frac{d \Psi}{ds}+B(x,y)\Psi=0 \eqno(6)
$$
where $\Psi(x,y)$ is two component vector $\Psi_1=z(s)$, $\Psi_2=t(s)$
and values $A(x,y)$ and $B(x,y)$ are the $2\times2$ matrix-functions.

   Note that full system of equations has the first integral
$$
2(za_3-\tau a_4){\dot x}^2+4(za_2-\tau a_3)\dot x \dot y+
2(za_1-\tau a_2){\dot y}^2+2 \dot x \dot z +2 \dot y \dot \tau=1,
$$
 equivalent to the relation
$$
z\dot x+t\dot y=\frac{s}{2}+\mu.
$$
This  allows to use only one linear second order differential equation
from the full matrix system (6) at the studying of the concrete examples.

  Thus, we have constructed the four-dimensional Riemannian space with the
metric (5) and with connection
$$
\Gamma_1=\left | \begin{array}{cccc}
-a_3 & -a_2 & 0 & 0 \\
a_4 & a_3 & 0 & 0 \\
z(a_{4y}- \alpha'')-\tau a_{4x} & z a_{3y}- \tau(a_{3x}+\alpha'') & a_3 & -a_4 \\
z(a_{3y}- 2 \alpha')-\tau (a_{3x}-\alpha'') & z(a_{2y}-\alpha)- \tau a_{2x} & a_2 & -a_3 \cr
\end{array} \right |,
$$
$$
\Gamma_2=\left |\begin{array}{cccc}
-a_2 & -a_1 & 0 & 0 \\
a_3 & a_2 & 0 & 0 \\
z a_{3y}- \tau(a_{3x}+\alpha'') & z(a_{2y}+\alpha)- \tau(a_{2x}+2\alpha') & a_2 & -a_3 \\
z(a_{2y}- \alpha)-\tau a_{2x} & z a_{1y}- \tau(a_{1x}+\alpha) & a_1 & -a_2
\end{array} \right |, \quad
$$
$$
\Gamma_3=\left |\begin{array}{cccc}
0 & 0 & 0 & 0 \\
0 & 0 & 0 & 0 \\
a_{3} & a_2 & 0 & 0 \\
a_{2} & a_{1} & 0 & 0
\end{array} \right |, \quad
\Gamma_4=\left | \begin{array} {cccc}
0 & 0 & 0 & 0 \\
0 & 0 & 0 & 0 \\
-a_{4} & -a_3 & 0 & 0 \\
-a_{3} & -a_{2} & 0 & 0
\end{array} \right |.
$$

    The curvature tensor of this metric has the form
$$
R^1_{112}=-R^3_{312}=-R^2_{212}=R^4_{412}= \alpha', \quad
R^1_{212}=-R^4_{312}= \alpha, \quad
R^2_{112}=-R^3_{412}=-\alpha'',
$$
$$
R^1_{312}= R^1_{412}= R^2_{312}=R^2_{412}=0,
$$
$$
R^3_{112}=2z(a_{2}\alpha''-a_{3}\alpha')+2\tau(a_4 \alpha'-a_3 \alpha''),
$$
$$
R^4_{212}=2z(a_{3}\alpha'-a_{2}\alpha)+2\tau(a_3 \alpha-a_2 \alpha'),
$$
$$
R^3_{212}=z(\alpha_x-\alpha'_y+ a_1\alpha''-a_3 \alpha)+\tau(\alpha''_y-
\alpha'_x+a_4\alpha-a_2\alpha''),
$$
$$
R^4_{112}=z(\alpha'_y-\alpha_x+a_1 \alpha''-a_3 \alpha)+\tau(\alpha'_x-
\alpha''_y+a_4\alpha-a_2\alpha'').
$$

   Using the expressions for components of projective curvature of
space $A_2$
$$
L_1= \alpha''_y- \alpha'_x+a_2 \alpha''+a_4 \alpha-2a_3 \alpha',
$$
$$
L_2= \alpha'_y- \alpha_x+a_1 \alpha''+a_3 \alpha-2a_2 \alpha',
$$
they  can be presented in form
$$
R^4_{112}=z(L_2+2a_2 \alpha'-2a_3 \alpha)-\tau(L_1+2a_3 \alpha' -
2a_4\alpha),
$$
$$
R^3_{212}=z(-L_2+2a_1 \alpha''-2a_2 \alpha')+\tau(L_1+2a_3 \alpha' -
-2a_2 \alpha''),
$$
$$
R_{13}=\left |\begin{array}{cccc}
0 & 0 & 0 & 0 \\
0 & 0 & 0 & 0 \\
0 & -\alpha' & 0 & 0 \\
\alpha' & 0 & 0 & 0
\end{array} \right |,\quad
R_{14}=\left |\begin{array}{cccc}
0 & 0 & 0 & 0 \\
0 & 0 & 0 & 0 \\
0 & \alpha'' & 0 & 0 \\
-\alpha'' & 0 & 0 & 0
\end{array} \right |, \quad
$$
$$
R_{23}=\left |\begin{array}{cccc}
0 & 0 & 0 & 0 \\
0 & 0 & 0 & 0 \\
0 & -\alpha & 0 & 0 \\
\alpha & 0 & 0 & 0
\end{array} \right | ,\quad
R_{24}=\left |\begin{array}{cccc}
0 & 0 & 0 & 0 \\
0 & 0 & 0 & 0 \\
0 & \alpha' & 0 & 0 \\
-\alpha' & 0 & 0 & 0
\end{array} \right |,
$$
$$
R^i_{j34}=0.
$$

     The Ricci tensor $R_{ik}=R^l_{ilk}$ of our space $D^4$ has the components
$$
R_{11}=2 \alpha'',\quad R_{12}=2 \alpha',\quad R_{22}=2 \alpha,
$$
and scalar curvature $R=g^{in}g^{km}R_{nm}$ of the space $D^4$ is
$R=0$.

Now we can introduce the tensor
$$
L_{ijk}=\nabla_k R_{ij}-\nabla_j R_{ik}=R_{ij;k}-R_{ik;j}.
$$
It has the following components
$$
L_{112}=-L_{121}=2L_1, \quad L_{221}=L_{212}=-2L_2
$$
and with help of them the invariant conditions which are
connected with the
 equations (1) may be constructed using the covariant derivations of the curvature tensor and the values
$L_1, L_2$ or with.

     The Weyl tensor of the space $D^4$ is
$$
C_{lijk}= R_{lijk}+\frac{1}{2}(g_{jl} R_{ik}+g_{ik} R_{jl}-
g_{jk} R_{il} - g_{il} R_{jk})+\frac{R}{6}(g_{jk} g_{il}- g_{jl} g_{ik}).
$$
It has only one component
$$
C_{1212}=tL_1-zL_2.
$$
Note that the values $L_1$ and $L_2$ in this formulae are the same
with the Liouville expressions in theory of invariants of the
equations (1).

   Using the components of the Riemann tensor
$$
R_{1412}=\alpha'',\quad R_{2412}=\alpha',\quad R_{2312}=-\alpha, R_{3112}=\alpha',
$$
$$
  R_{1212}=z(\alpha_x-\alpha'_y+a_1\alpha''-2a_2\alpha'+a_3\alpha)+
t(\alpha''_y-\alpha'_x+a_4\alpha-2a_3\alpha'-a_2\alpha'')
$$
the equation
$$
\left |\begin{array}{c}
 R_{AB}-\lambda g_{AB}
\end{array} \right |=0
$$
for determination of the  Petrov type of  the spaces $D^4$ have been considered.
 Here $R_{AB}$ is symmetric $6\times6$ matrix constructed
from the components of the Riemann tensor $R_{ijkl}$ of the space $D^4$.

  In particular we have checked that all scalar invariants of the space $D^4$
of this sort
$$
R_{ij}R^{ij}=0,\quad R_{ijkl}R^{ijkl}=0, ...
$$
constructed from the curvature tensor of the space $M^4$ and its covariant
derivations  are equal to zero.
\begin{rem}

    The spaces with metrics (5) are flat for the equations (1) with the
conditions
$$
\alpha=0, \quad \alpha'=0, \quad \alpha''=0,
$$
on coefficients $a_i(x,y)$.

Such type of equations have the components of projective curvature
$$
L_1=0, \quad L_2=0
$$
and they are reduced to the the form $y''=0$ with help of the points
transformations.

   On the other hand there are examples of equations (1) with conditions
$L_1=0, \quad L_2=0$ but
$$
\alpha \neq 0, \quad \alpha'\neq 0, \quad \alpha''\neq0.
$$
For such type of equations the curvature of corresponding Riemann
spaces is not equal to zero.

   In fact, the equation
$$
y''+2e^{\varphi}y'^3-\varphi_y y'^2+\varphi_x y'-2e^{\varphi}=0 \eqno(7)
$$
where the function $\varphi(x,y)$ is solution of the
Wilczynski-Tzitzeika nonlinear equation integrable by the Inverse
Transform Method.
$$
\varphi_{xy}=4e^{2 \varphi}-e^{-\varphi}. \eqno(7')
$$
has conditions  $L_1=0, \quad L_2=0$ but
$$
\alpha \neq 0, \quad \alpha'\neq 0, \quad \alpha''\neq0.
$$
In particular even for the linear second order dfferential equations we
have a non flat Riamannian spaces.
\end{rem}
\begin{rem}

     The studying of the properties of the Riemann spaces with the metrics (5)
for the equations (2) with chaotical behavior at the values of
coefficients $(\sigma=10,\quad b=8/3,\quad r > 24)$ is important
problem. The spaces with such values of parameters have specifical
relations between the components of curvature tensor.

   To studying this problem the geodesic deviation equation
$$
\frac{d^2 \eta^i}{ds^2}+2\Gamma^{i}_{lm}\frac{dx^m}{ds}\frac{d\eta^l}{ds}+
\frac{\partial \Gamma^{i}_{kl}}{\partial x^j}\frac{dx^k}{ds}
\frac{dx^l}{ds}\eta^j=0
$$
where $\Gamma^i_{lm}$ are the Christoffell coefficients of the
metrics (5) with the  coefficients
$$
a_1=0,\quad a_2=-\frac{1}{y}, \quad a_3=(\frac{\alpha y}{3}-\frac{1}{3x}),
$$
$$
a_4=\epsilon xy^4-\beta x^3 y^4-
\beta x^2y^3-\gamma y^3 + \delta \frac{y^2}{x}.
$$
may be used.

   For the equations
$$ y''+a_4(x,y)=0 $$ the four-dimensional Riemann spaces with the
metrics $$ ds^2=-2t a_4 dx^2 +2dxdz+2dydt
 $$
  and geodesics in form
$$ \ddot x=0, \quad \ddot y+a_4(x,y)(\dot x)^2=0,\quad \ddot
t+a_{4y}(\dot x)^2 t=0 $$ $$ \ddot z-t a_{4x}(\dot x)^2-2t
a_{4y}\dot x \dot y- 2a_4 \dot x \dot t=0 $$ are connected.

   It is interesting to note that in case of the Painleve II equation
$$
y''=2y^3+xy+\alpha
$$
the system for geodesic deviations of the corresponding Riemann
space
$$
\frac{d^2 \eta^1}{d s^2}=0,\quad \frac{d^2 \eta^2}{d s^2}=(6y^2+s)\eta^2+4y^3+3sy+\alpha,
$$
$$
\frac{d^2 \eta^3}{d s^2}=-(12ty^2+2ts)\frac{d \eta^3}{d
s}-(4y^3+2sy+2\alpha) \frac{d \eta^4}{d s}-(t+24ty\dot y+2y^2\dot
t+2s\dot)\eta^2-
$$
$$
(y+12y^2\dot y+2s\dot y)\eta^4-2ty-4ts\dot y-12ts^2\dot y-4y^3\dot t-
4sy\dot t-2\alpha\dot t
$$
$$
\frac{d^2 \eta^4}{d s^2}=(6y^2+s)\eta^4+12ty^2+3ts+12ty\eta^2.
$$
depends on the parameter $\alpha$.

    For the  equations
$$
y''+3a_3(x,y)y'+a_4(x,y)=0
$$
  a corresponding Riemann spaces have the metrics
$$
 ds^2=2(za_3-\tau a_4)dx^2-4 \tau a_3dxdy+2dxdz+2dyd \tau
$$ and
the equations of geodesics $$ \ddot x -a_3\dot x^2=0,\quad \ddot
y+2a_3 \dot x \dot y+a_4\dot x^2=0, $$ $$ \ddot \tau -2a_3 \dot x
\dot \tau-a_{3y}\dot x^2 z+(a_{4y}-2{a_3}^2- 2a_{3x})\dot x^2 \tau
=0, $$ $$ \ddot z +2a_3 \dot x \dot z-2(a_3 \dot y +a_4 \dot x)
\dot \tau+ [(a_{3x}+2{a_3}^2)\dot x^2+2a_{3y}\dot x \dot y]
z-[a_{4x}\dot x^2+2(a_{4y}- 2a_3^2)\dot x \dot y+2a_{3y} \dot y^2]
\tau=0. $$
\end{rem}

     Let us consider  the possibility of embedding of
the spaces with the metrics (5) using the facts from the theory of
embedding of Riemann spaces into the spaces with a flat metrics.

      For the Riemann spaces of the class one (which can be embedded into
the 5-dimensional Euclidean space) the following conditions are
fulfilled
$$
R_{ijkl}=b_{ik}b_{jl}-b_{il}b_{jk}
$$
and
$$
b_{ij;k}-b_{ik;j}=0
$$
where $R_{ijkl}$ are the components of curvuture tensor of the space with
metrics $ds^2=g_{ij}dx^idx^j$.

  The consideration of these relations for the spaces with the metrics (5)
 lead to  the conditions on the values $a_i(x,y)$ from which follows
 that the embedding in the fife-dimensional space with the flat metrics
 is possible only in case
$$
a_i(x,y)=0.
$$

    For the spaces of the class two (which admits the embedding into
the 6-dim Eucledean space with some signature)
the conditions for that  are more complicated. They are
$$
R_{abcd}=e_1(\omega_{ac}\omega_{bd}-\omega_{ad}\omega_{bc})+
e_2(\lambda_{ac}\lambda_{bd}-\lambda_{ad}\lambda_{bc}),
$$
$$
\omega_{ab;c}-\omega_{ac;b}=e_2(t_c \lambda_{ab}-t_b \lambda_{ac}),
$$
$$
\lambda_{ab;c}-\lambda_{ac;b}=-e_1(t_c \omega{ab}-t_b \omega_{ac}),
$$
$$
t_{a;b}-t_{a;c}=\omega_{ac}\lambda^c_b-\lambda_{ac} \omega^c_b.
$$
and lead to the relations
$$
\epsilon^{abcd}\epsilon^{nmrs}\epsilon^{pqik}R_{abnm}R_{cdpq}R_{rsik}=0
$$
and
$$
\epsilon^{cdmn} R_{abcd}R^{ab}_{mn}=-8e_1e_2\epsilon^{cdmn}t_{c;d}t_{m;n}
$$

\section{On relation with theory of the surfaces}

    The existence of the Riemann metrics for the equations (1) may be
used for construction of the corresponding surfaces.

    One possibility  concerns the study of two-dimensional subspaces
of a given 4-dimensional space  which are the generalization of the
surfaces of translation. The equations for coordinates $Z^{i}(u,v)$ of
such type of the surfaces are
$$
 \frac{\partial^2 Z^{i}}{\partial u \partial v} +
\Gamma^{i}_{jk} \frac{\partial Z^{j}}{\partial u}
 \frac{\partial Z^{k}}{\partial v} =0.   \eqno(8)
$$
where $\Gamma_i^{jk}$ are the components of connections.

       Let us consider the system (8) in detail.

   We get the following system of equations for coordinates $x=x(u,v),
\quad y=y(u,v),\quad z=z(u,v),\quad t=t(u,v)$
$$
x_{uv}-a_3 x_u x_v-a_2(x_u y_v+ x_v y_u)-a_1 y_u y_v=0,
$$
$$
y_{uv}+a_4x_u x_v+a_3(x_u y_v+x_v y_u)+a_2 y_u y_v=0,
$$
$$
z_{uv}+ z_u[a_2 y_v+a_3 x_ v]+ z_v[a_2 y_u+a_3 x_u]-t_u[a_3 y_v+a_4 x_v]-
t_v[a_4 x_u+a_3 y_u]+z[2y_u y_v a_1 a_3-y_v y_u a_{1x}+x_v x_u a_{3x}-
$$
$$
2x_u x_v a_2 a_4+y_u x_v a_{3y}-y_v x_u a_{3y}-2y_v y_u (a_2)^2+
2y_v y_u a_{2y}+
2 x_v y_u (a_3)^2]+t[y_v y_u a_{2x}- x_v x_u a_{4_x}- x_v y_u a_{4y}+
$$
$$
2y_u y_v a_2 a_3-2 y_v x_u a_2 a_4-2y_v y_u a_{3y}+2x_v y_u (a_3)^2+
2y_v x_u (a_3)^2- y_v x_u a_{4y}-2y_u x_v a_2 a_4-2 y_v y_u a_1 a_4]=0
$$

$$
t_{uv}+ z_u[a_2 x_v+a_1 y_v]+z_v[a_2 x_u+a_1 y_u]-t_u[a_3 x_v+a_2 y_v]-
t_v[a_3 x_u+a_2 y_u]+z[-2x_u y_v a_1 a_3+x_v y_u a_{1x}+2 x_v x_u a_{2x}+
$$
$$
2x_u x_v a_2 a3+y_v y_u (a_{1y}+2 y_v x_u (a_2)^2-2x_v x_u a_4
a_1- x_v x_u a_{3y}+2 x_v y_u (a_2)^2- 2 x_v y_u a_1a_3+y_v x_u
a_{1_x}]+
$$
$$
t[-y_v x_u a_{2x}-x_v y_u a_{2x}+x_v x_u a_{4y}+ 2 x_u x_v a_2a_4-
2y_u y_v a_1a_3- 2 x_v x_u a_{3x}- 2 x_v x_u (a_3)^2+ 2y_v y_u
(a_2)^2- y_v y_u a_{2y}]=0
$$

   Here were used the expressions for the Christoffell coefficients
$$
\Gamma^1_{11}=-a_3(x,y),\quad \Gamma^2_{11}=a_4(x,y),\quad \Gamma^1_{12}=-a_2(x,y),
\quad \Gamma^2_{12}=a_3(x,y),
$$
$$
\Gamma^3_{13}=a_3(x,y),\quad \Gamma^4_{13}=a_2(x,y),\quad\Gamma^3_{14}=-a_4(x,y),
\quad\Gamma^4_{14}=-a_3(x,y),
$$
$$
\Gamma^1_{22}=-a_1(x,y),\quad \Gamma^2_{22}=-a_2(x,y),\quad \Gamma^3_{23}=a_2(x,y),
$$
$$
\Gamma^4_{23}=a_1(x,y),\quad\Gamma^3_{24}=-a_3(x,y),\quad\Gamma^4_{24}=-a_2(x,y),
$$
$$
\Gamma^3_{11}=z\frac{\partial a_3(x,y)}{\partial x}-t\frac{\partial a_4(x,y)}{\partial x}+
2 z a_3(x,y)^2-2z a_2(x,y)a_4(x,y),
$$
$$
\Gamma^4_{11}=2z\frac{\partial a_2(x,y)}{\partial x}-2t\frac{\partial a_3(x,y)}{\partial x}-
z\frac{\partial a_3(x,y)}{\partial y}+
$$
$$
t\frac{\partial a_4(x,y)}{\partial y}+
2 z a_3(x,y)a_2(x,y)-2 z a_1(x,y)a_4(x,y)+2t a_2(x,y)a_4(x,y)-2t a_3(x,y)^2,
$$
$$
\Gamma^3_{12}=z\frac{\partial a_3(x,y)}{\partial y}-t\frac{\partial a_4(x,y)}{\partial y}+
2 t a_3(x,y)^2-2t a_2(x,y)a_4(x,y),
$$
$$
\Gamma^4_{12}=z\frac{\partial a_1(x,y)}{\partial x}-t\frac{\partial a_2(x,y)}{\partial x}+
2 z a_2(x,y)^2-2z a_1(x,y)a_3(x,y),
$$
$$
\Gamma^4_{22}=z\frac{\partial a_1(x,y)}{\partial y}-t\frac{\partial a_2(x,y)}{\partial y}+
2 t a_3(x,y)^2-2t a_1(x,y)a_3(x,y),
$$
$$
\Gamma^3_{22}=2z\frac{\partial a_2(x,y)}{\partial y}-2t\frac{\partial a_3(x,y)}{\partial y}-
z\frac{\partial a_1(x,y)}{\partial x}+
$$
$$
t\frac{\partial a_2(x,y)}{\partial x}+
2 z a_3(x,y)a_1(x,y)-2 t a_1(x,y)a_4(x,y)+2t a_2(x,y)a_3(x,y)-2z a_2(x,y)^2.
$$

    From these relations we can see that two last equations of the full system
are linear and have the form of the linear $2\times2$ matrix Laplace equations
$$
\frac{\partial^2 \Psi}{\partial u \partial v}+A\frac{\partial \Psi}{\partial u}+
B\frac{\partial \Psi}{\partial v}+C \Psi=0.     \eqno(9)
$$
We can integrate them with the help of generalization of the
 Laplace-transformation [26].

   For that we use the transformations
$$
\Psi_1=(\partial_v+A)\Psi,\quad (\partial_u+B)\Psi_1=h\Psi
$$
where the Laplace invariants are
$$
H=A_u+BA-C,\quad K=B_v+AB-C
$$
and then construct a new equation of type (7) for the function
$\Psi_1$ with a new invariants
$$
A_1=HAH^{-1}-H_vH^{-1},\quad B_1=B,\quad C_1=B_v-H+(HAH^{-1}-H_vH^{-1})B,
$$

     Let us consider some examples.

    The first example concerns the conditions
$$
x=x,\quad y=y,\quad u=x,\quad v=y,\quad z=z(u,v)=z(x,y),\quad t=t(x,y)=t(u,v).
$$

     From the first equations of the full system we get
$$
a_2(x,y)=0,\quad a_3(x,y)=0
$$
and from next two we have the system of equations
$$
\frac{\partial^2 z}{\partial x \partial y}- \frac{\partial a_4(x,y)}{\partial y}t-
a_4(x,y)\frac{\partial t}{\partial y}=0,
$$
$$
\frac{\partial^2 t}{\partial x \partial y}+ \frac{\partial a_1(x,y)}{\partial x}z+
a_1(x,y)\frac{\partial z}{\partial x}=0,
$$

   They are equivalent to the independent relations
$$
\frac{\partial z}{\partial x}-t a_4(x,y)=0,
$$
$$
\frac{\partial t}{\partial y}+z a_1(x,y)=0,
$$
or
$$
\frac{\partial^2 z}{\partial x \partial y}-
\frac{1}{a_4}\frac{\partial a_4(x,y)}{\partial y}\frac{\partial z}{\partial x}+
a_1(x,y)a_4(x,y) z=0,
$$
$$
\frac{\partial^2 t}{\partial x \partial y}-
\frac{1}{a_1}\frac{\partial a_1(x,y)}{\partial x}\frac{\partial t}{\partial y}+
a_1(x,y)a_4(x,y) t=0.
$$

  Any solution of this system of equations give us the examples of the surfaces which
corresponds to the second order ODE's in form
$$
\frac{d^2 y}{d x^2}+a_1(x,y)\left(\frac{d y}{d x}\right)^3+a_4(x,y)=0
$$

    Next example concerns the conditions:
$$
x=u+v,\quad y=uv.
$$

   From the first relations we get the system for the coefficients
$a_i(x,y)$
$$
a_3+xa_2=-ya_1,\quad ya_2+xa_3=-1-a_4
$$
from which we derive the expressions
$$
a_2=\frac{1+a_4(x,y)-xya_1(x,y)}{x^2-y},\quad a_3=\frac{y^2a_1(x,y)
-x-xa_4(x,y)}{x^2-y}.
$$

As result we get the equations
$$
\frac{d^2 y}{d x^2}+a_1\left(\frac{d y}{d x}\right)^3+
3\frac{(1+a_4-xya_1)}{x^2-y}\left(\frac{d y}{d x}\right)^2+
3\frac{(y^2a_1-x-xa_4)}{x^2-y}\frac{d y}{d x}+a_4=0
$$

   In particular case
$$
a_1(x,y)=0,\quad a_4(x,y)=\frac{-x^2}{y}
$$
we get the equation
$$
\frac{d^2 y}{d x^2}-\frac{3}{y}\left(\frac{d y}{d x}\right)^2+
\frac{3x}{y}\frac{d y}{d x}-\frac{x^2}{y}=0    \eqno(10)
$$
and the equations for the coordinates of the corresponding
surfaces
$$
t_{uv}-\frac{t_u}{u}-\frac{t_v}{v}-\frac{z_u}{uv}-\frac{z_v}{uv}+
\frac{u+v}{u^2v^2}z+\frac{uv+u^2+v^2}{u^2v^2}t=0,
$$
$$
z_{uv}+\frac{z_u}{u}+\frac{z_v}{v}+\frac{u+v}{v}t_v+\frac{u+v}{u}t_u-
\frac{uv+u^2+v^2}{u^2v^2}z-\frac{u^2v+u^3+v^3+v^2u}{u^2v^2}t=0.
$$

     Note that the equation (10) can be transformed to the form
$$
\frac{d^2 z}{d \rho^2}-\frac{3}{z}\left(\frac{d z}{d \rho}\right)^2+
\left(\frac{3}{z}-9\right)\frac{d z}{d \rho}-10z+6-\frac{1}{z}=0    \eqno(11)
$$
with help of the substitution
$$
y(x)=x^2z(\ln(x)).
$$

      Another possibility for the studying of two-dimensional surfaces
in space with metrics (5) concerns the choice of section
$$
x=x, \quad y=y, \quad z=z(x,y),\quad \tau=\tau(x,y)
$$
in space with the metrics (5).

    Using the expressions
$$
dz=z_x dx+z_y dy, \quad d\tau=\tau_x dx+ \tau_y dy
$$
we get the metric
$$
ds^2=2(z_x+za_3 - \tau a_4)dx^2 +2(\tau_x + z_y +2za_2 - 2 \tau a_3)dxdy+
2(\tau_y + za_1- \tau a_2)dy^2.
$$

    We can use this presentation for investigation of particular cases of
equations (1).

 1. The choice of the functions $z,\quad \tau$ in form
$$
z_x+za_3 - \tau a_4=0,
 $$
$$
 \tau_x + z_y +2za_2 - 2 \tau a_3=0       \eqno (12)
$$
 $$
  \tau_y + za_1- \tau a_2=0
  $$
 is connected with a flat surfaces
and is reduced at the substitution $$ z=\Phi_x, \quad \tau=\Phi_y
$$ to the system
 $$
 \Phi_{xx}=a_4\Phi_y-a_3\Phi_x,
$$
$$
\Phi_{xy}=a_3\Phi_y-a_2\Phi_x,
$$
$$
\Phi_{yy}=a_2\Phi_y-a_1\Phi_x.
$$
compatible at the conditions
$$
\alpha=0, \quad \alpha'= 0, \quad \alpha''=0.
$$
\begin{rem}
      The choice of the functions $z=\Phi_x, \tau=\Phi_y$
satisfying the system of equations
$$
 \Phi_{xx}=a_4\Phi_y-a_3\Phi_x,
$$
$$
\Phi_{yy}=a_2\Phi_y-a_1\Phi_x
$$
with the coefficients $a_i(x,y)$ in form
$$
a_4=R_{xxx}, \quad a_3=-R_{xyy}, \quad a_2=R_{xyy}, \quad a_1=R_{yyy}
$$
where the function $R(x,y)$ is the solution of WDVV-equation
$$
R_{xxx}R_{yyy}-R_{xxy}R_{xyy}=1
$$
correspond to the equations (1)
$$
y''-R_{yyy}y'^{3}+3R_{xyy}y'^{2}-3R_{xxy}y'+R_{xxx}=0.
$$
\end{rem}
      The following choice of the coefficients $a_i$
$$
a_4=-2\omega,\quad a_1=2\omega,\quad a_3=\frac{\omega_x}{\omega},\quad
a_2=-\frac{\omega_y}{\omega}
$$
lead to the system
$$
 \Phi_{xx}+\frac{\omega_x}{\omega}\Phi_x+2\omega\Phi_y=0,
$$
$$
 \Phi_{yy}+2\omega\Phi_x+\frac{\omega_y}{\omega}\Phi_y=0
$$ with condition of compatibility $$ \frac{\partial^2 \ln
\omega}{\partial x \partial y}= 4\omega^2+\frac{\kappa}{\omega} $$
which is the  Wilczynski-Tzitzeika-equation.

\begin{rem}

    The  linear system of equations for the WDVV-equation  some surfaces
in 3-dim projective space is determined. In  canonical form it
becomes [13] $$
 \Phi_{xx}-R_{xxx} \Phi_y+(\frac{R_{xxxy}}{2}-
\frac{R_{xxy}^2}{4}-\frac{R_{xxx} R_{xxy}}{2})\Phi=0,
$$
$$
 \Phi_{yy}-R_{yyy} \Phi_x+(\frac{R_{yyyx}}{2}-
\frac{R_{xyy}^2}{4}-\frac{R_{yyy} R_{xxy}}{2})\Phi=0,
$$

     The relations between invariants of Wilczynsky for the linear system
is correspondent to the various types of surfaces. Some of them
with the solutions of WDVV equation are connected.

\end{rem}
\begin{rem}
From the elementary point of view the surfaces which are connected
with the system of equations like the Lorenz can be constructed in
such a way.

   From the assumption
$$
z=z(x,y)
$$
we get
$$
\sigma(y-x)z_x+(rx-y-zx)z_y=xy-bz.
$$
The solutions of this equation give us the examples of the surfaces
$z=z(x,y)$.

 The Riemann metrics of the space which connected with the equation
(2) has the form
$$
 ds^2=(\frac{2}{3}\alpha z y-\frac{2}{3x} z-
\frac{2}{x}\delta y^2 t-2\epsilon t x y^4 +2\beta t x^3 y^4+2\beta
t x^2 y^3+2\gamma t y^3)dx^2+
$$
$$
2(-2 \frac{z}{y}-\frac{2}{3}\alpha t
y+\frac{2}{3 x}t)dxdy+\frac{2}{y}t dy^2+2dx dz+2 dy dt.
$$

   The properties of the space with a such metrics from the parameters $\alpha$, $\beta$,
$\gamma$ $\delta$, $\epsilon$ are determined and may be very
specifical when the Lorenz dynamical system has the strange attractor.
\end{rem}

\section{ Symmetry, the Laplace-Beltrami equation, tetradic presentation}

     Let us consider the system of equations
$$
\xi_{i,j}+\xi_{j,i}=2\Gamma^k_{ij} \xi_k
$$
for the Killing vectors of metrics (5).
It has the form
$$
\xi_{1x}=-a_3 \xi_1+a_4 \xi_2+(zA-t a_{4x})\xi_3+(zE+tF)\xi_4,
$$
$$
\xi_{2y}=-a_1 \xi_1+a_2 \xi_2+(zC+tD)\xi_3+(za_{1y}-tH)\xi_4,
$$
$$
\xi_{1y}+\xi_{2x}=2[-a_2 \xi_1+a_3 \xi_2+(za_{3y}-tB)\xi_3+(zG-ta_{2x})\xi_4,
$$
$$
\xi_{1z}+\xi_{3x}=2[a_3 \xi_3+a_2 \xi_4],\quad
\xi_{1t}+\xi_{4x}=2[-a_4 \xi_3-a_3 \xi_4],
$$
$$
\xi_{2z}+\xi_{3y}=2[a_2 \xi_3+a_1 \xi_4],\quad
\xi_{2t}+\xi_{4y}=-2[a_3 \xi_3-a_2 \xi_4],
$$
$$
\xi_{3z}=0,\quad \xi_{4t}=0.
$$

    In particular case $\xi_i(x,y)$
$$
\xi_3 = \xi_4 =0, \quad \xi_i=\xi_i(x,y)
$$
we get the system of equations
$$
\xi_{1x}=-a_3 \xi_1+a_4 \xi_2,\quad \xi_{2y}=-a_1 \xi_1+a_2 \xi_2,
$$
$$
\xi_{1y}+\xi_{2x}=2[-a_2 \xi_1+a_3 \xi_2]
$$
equivalent to the system for the z=z(x,y) and $\tau=\tau(x,y)$
same with the system (12), connected with integrable equations.

By analogy the system of equations for the Killing tensor
$$
K_{ij;l}+K_{jl;i}+K_{li;j}=0
$$
and the Killing-Yano tensor $Y_{ij}+Y_{ji}=0$
$$
Y_{ij;l}+Y_{il;i}=0
$$
 may be investigated.

\begin{rem}
      The Laplace-Beltrami operator
$$
\Delta= g^{ij}(\frac{\partial^2}{\partial x^i \partial x^j}-
\Gamma^k_{ij}\frac{\partial}{\partial x^k})
$$
can be used for investigation of the properties of the metrics (5).

   For example the equation
$$
\Delta \Psi=0
$$
has the form
$$
(ta_4-za_3)\Psi_{zz}+2(ta_3-za_2)\Psi_{zt}+(ta_2-za_1)\Psi_{tt}+\Psi_{xz}+
\Psi_{yt}=0.
$$

     Some solutions of this equation with geometry of the metrics
(5) are connected.

     Putting the expression
$$
\Psi= \exp[zA+tB]
$$
into the equation
$$
\Delta\Psi=0
$$
we get the conditions
$$
A=\Phi_y,\quad B=-\Phi_x,
$$
and
$$
a_4 \Phi^2_y-2a_3 \Phi_x \Phi_y+
a_2 \Phi^2_x-\Phi_y \Phi_{xx}+\Phi_x \Phi_{xy}=0,
$$
$$
a_3 \Phi^2_y-2a_2\Phi_x\Phi_y+
a_1\Phi^2_x-\Phi_y \Phi_{xy}+\Phi_x\Phi_{yy}=0,
$$

   Another possibility for the studying of the properties of
a given Riemann spaces is connected with computation of the heat
invariants of the Laplace-Beltrami operator.

    For that the fundumental solution $K(\tau,x,y$ of the heat equation
$$
\frac{\partial \Psi}{\partial \tau}=
g^{ij}(\frac{\partial^2\Psi}{\partial x^i \partial x^j}-
\Gamma^k_{ij}\frac{\partial \Psi}{\partial x^k})
$$
is considered.

   The function $K(\tau,x,y)$ has the following
asymptotic expansion on diagonal as $t\rightarrow 0+$
$$
K(\tau,x,x)=\sim\sum_{n=0}^\infty a_n(x)\tau^{n-2}
$$
and the coefficients $a_n(x)$ are local invariants (heat invariants)
of the Riemann space $D^4$ with the metrics (5).

   In turn the eikonal equation
$$
 g^{ij} \frac{\partial F}{\partial x^i} \frac{\partial F}{\partial x^j}=0
$$
or
$$
F_xF_z+F_yF_t-(ta_4-za_3)F_zF_z-2(ta_3-za_2)F_zF_t-(ta_2-za_1)F_tF_t=0.
$$
also can be used for investigation of the properties of isotropical
surfaces in the space with metrics (5).

     In particular case the solutions of eikonal equation in form
$$
F=A(x,y)z^2+B(x,y)zt+C(x,y)t^2+D(x,y)z+E(x,y)t
$$
lead to the following conditions on coefficients
$$
2AA_x + BA_y - a_1 B^2 - 4a_2 AB - 4a_3 A^2=0,
$$
$$
2AB_x + BA_x + 2CA_y + BB_y - 4a_1 BC - a_2(B^2+8AC) + 4a_4 A^2=0,
$$
$$
2CB_y + BC_y + 2AC_x + BB_x - 4a_1 C^2 + a_3(B^2+8AC) + 4a_4 AB =0,
$$
$$
2CC_y + BC_x + 4a_2 C^2 + 4a_3 BC + a_4 B^2=0,
$$
$$
2AD_x + DA_x + EA_y + BD_y - 2a_1 BE - 2a_2(BD+2AE) - 4a_3 AD =0,
$$
$$
2CD_y + (BD)_x + 2AE_x + (BE)_y - 4a_1 EC - 4a_2 CD + 4a_3 AE + 4a_4 AD =0,
$$
$$
2CE_y + CE_y + DC_x + BE_x - 4a_2 CE + 2a_3(BE+2CD) + 2a_4 BD =0,
$$
$$
DD_x + ED_y -a_1 E^2 - 2a_2 DE - a_3 D^2=0,
$$
$$
EE_y + DE_x +a_2 E^2 + 2a_3 DE + a_4 D^2=0
$$
which may be used for the theory of equations (1).
\end{rem}
\begin{rem}
   The metric (5) has a tetradic presentation
$$
g_{ij}=\omega_{i}^a \omega_{j}^b \eta_{ab}
$$
where
$$
\eta_{ab}=\left |\begin{array}{cccc}
0 & 0 & 1 & 0 \\
0 & 0 & 0 & 1 \\
1 & 0 & 0 & 0 \\
0 & 1 & 0 & 0
\end{array} \right |.
$$

      For example we get
$$
ds^2= 2 \omega^1 \omega^3 +2\omega^2 \omega^4
$$
where
$$
\omega^1=dx+dy,\quad \omega^2=dx+dy+\frac{1}{t(a_2-a_4)}(dz-dt),
$$
$$
\omega^4=-t(a_4dx+a_2dy),\quad \omega^3=z(a_3dx+a_1dy)+\frac{1}{(a_2-a_4)}
(a_2dz-a_4dt).
$$
and
$$
a_1+a_3=2a_2, \quad  a_2+a_4=2a_3.
$$
\end{rem}
\begin{rem}
   Some of equations on curvature tensors in space $M^4$ are connected with
ODE's. For example, the equation
$$
R_{ij;k}+R_{jk;i}+R_{ki;j}=0
$$
lead to the conditions on coefficients $a_i(x,y)$
$$
\alpha''_x+2a_3 \alpha''-2a_4 \alpha'=0,
$$
$$
\alpha_y+2a_1 \alpha'-2a_2 \alpha=0,
$$
$$
\alpha''_y+2 \alpha'_x+4a_2 \alpha''-2a_4 \alpha-2a_3 \alpha'=0,
$$
$$
\alpha_x+2 \alpha'_y-4a_3 \alpha +2a_2 \alpha'+2a_1 \alpha''=0.
$$

     The solutions of this system give us  the second
order equations connected with the space $D^4$ with a given
condition on the Ricci tensor. The simplest examples are
$$
y''-\frac{3}{2y}y'^2+y^3=0, \quad y''-\frac{3}{y}y'^2+y^4=0,\quad
y''+3(2+y)y'+y^3+6y^2-16=0.
$$

   It is of interest to note that the above system is the same with the
 Liouville system for geodesics from the Proposition 1.

     The studying of the invariant conditions like
$$
R_{ij;k}-R_{jk;i}=R^{n}_{ijk;n},\quad \Box R_{ijkl}=0,\quad \Box
R_{ijkl;m}=0
$$
is also interesting for theory of equations (1).
\end{rem}

\begin{rem}

    The construction of the Riemannian extension of
two-dimensional spaces  connected with ODE's of type (1)
can be generalized for three-dimensional case with the equations
of the form
$$
\ddot x+A_1(\dot x)^2+2A_2\dot x\dot y+2A_3\dot x\dot z+
A_4(\dot y)^2+2A_5\dot y \dot z+A_6(\dot z)^2=0,
$$
$$
\ddot y+B_1(\dot x)^2+2B_2\dot x\dot y+2B_3\dot x\dot z+
B_4(\dot y)^2+2B_5\dot y \dot z+B_6(\dot z)^2=0,
$$
$$
\ddot z+C_1(\dot x)^2+2C_2\dot x\dot y+2C_3\dot x\dot z+
C_4(\dot y)^2+2C_5\dot y \dot z+C_6(\dot z)^2=0.
$$
or
$$
y''+c_0 +c_1x'+c_2y'+c_3{x'}^2+c_4x'y'+c_5{y'}^2+
y'(b_0+b_1x'+b_2y'+b_3{x'}^2+b_4x'y'+
b_5{y'}^2)=0,
$$
$$
x''+a_0 +a_1x'+a_2y'+a_3{x'}^2+a_4x'y'+a_5{y'}^2+
x'(b_0+b_1x'+b_2y'+b_3{x'}^2+b_4x'y'+
b_5{y'}^2)=0,
$$
where $a_i, b_i, c_i$ are the functions of variables $x,y,z$.

    Corresponding expression for the 6-dimensional metrics are:
$$
ds^2=-2(A_1 u+B_1 v+C_1 w)dx^2-4(A_2u+B_2v+C_2w)dxdy-
4(A_3u+B_3v+C_3w)dxdz-
$$
$$
2(A_4u+B_4v+C_4w)dy^2-4(A_5u+B_5v+C_5w)dydz-2(A_6u+B_6v+C_6w)dz^2+2dxdu+
2dydv+2dzdw
$$

   This gives us the possibility to study the properties of such type of
 equations from geometrical point of view.

   Let us consider some examples.
\end{rem}

\section{The Riemann metrics of zero curvature and the KdV equation}

     The system of matrix equations in form
$$
\frac{\partial\Gamma_2}{\partial x}-\frac{\partial\Gamma_1}{\partial y}+
[\Gamma_1,\Gamma_2]=0,
$$
$$
\frac{\partial\Gamma_3}{\partial x}-\frac{\partial\Gamma_1}{\partial z}+
[\Gamma_1,\Gamma_3]=0,  \eqno(13)
$$
$$
\frac{\partial\Gamma_3}{\partial y}-\frac{\partial\Gamma_2}{\partial z}+
[\Gamma_2,\Gamma_3]=0,
$$
where $\Gamma_i(x,y,z)$-are the $3 \times 3$ matrix functions with conditions
$\Gamma^k_{ij}=\Gamma^k_{ji}$ are considered.

  This system can be considered as the condition of the zero curvature
of the some $3-dim$ space equipping by the affine connection with
coefficients $\Gamma(x,y,z)$.

     Let $\Gamma^k_{ij}(x,y,z)$ be in the form
$$
\Gamma_1=y^2B_1(x,z)+yA_1(x,z)+C_1(x,z)+\frac{1}{y}D_1(x,z),
$$
$$
\Gamma_3=y^2B_3(x,z)+yA_3(x,z)+C_3(x,z)+\frac{1}{y}D_3(x,z)+\frac{1}{y^2}E_3(x,z)
$$
$$
\Gamma_2=C_2(x,z)+\frac{1}{y}D_2(x,z).
$$

   Then after substituting these expressions in formulas (13) we get
the system of nonlinear equations for components of affine
connection. Some of these equations may be of interest for
applications.

    Let us consider the space with the metrics
$$
g_{ik}=\left(\begin{array}{ccc}
y^2 & 0 & y^2 l(x,z)+m(x,z)\\
0 & 0 & 1\\
 y^2 l(x,z)+m(x,z)& 1&y^2l(x,z)^2-2 y l_x(x,z)+2l(x,z)m(x,z)+2n(x,z)\\
\end{array} \right ).
$$

   Using the relations between the metrics and connection
$$
\Gamma^k_{ij}=\frac{1}{2}g^{kl}(\partial_ig_{jl}+\partial_jg_{il}-
\partial_lg_{ij})
$$
we get the components of matrices $\Gamma_i$
$$
\Gamma_1=\left(\begin{array}{ccc}
yl+m/y & 1/y &yl^2+ml/y\\
-(y^2l_x-2yn+m^2/y-m_x)&-m/y &-(y^2ll_x-2yln+
lm^2/y+yl_{xx}-ml-lm_x -n_x \\
-y  & 0  &-ly
\end{array} \right ),
$$
$$
\Gamma_2=\left(\begin{array}{ccc}
1/y & 0 & l/y\\
-m/y & 0 &-(lm/y+l_x)\\
0 & 0 & 0
\end{array} \right ),
$$
$$
\Gamma_3=\left(\begin{array}{lll}
l^2+m/y&1/y&l_z+m_z/y^2-2ll_x+l_{xx}/y-2ml_x/y^2-\\
& & lm_x/y^2-n_x/y^2+yl^3+l^2m/y-\\
(y^2ll_x-2lny+lm^2/y+yl_{xx}-\\& & \\
 ml_x-lm_x-n_x )& -l_x-lm/y&
-(y^2l^2l_x+yll_{xx}-2lml_x-l^2m_x- \\& &ln_x+mm_z/y^2+ ml_{xx}/y-
2m^2l_x/y^2-\\& &lmm_x/y^2- mn_x/y^2-2yl_x^{2}-\\& &
2ynl^2+2nl_x+m^2l^2/y+yl_{zx}-n_z)\\
-yl & 0 &-yl^2+l_x
\end{array} \right ),
$$

  In the case $l(x,z)=n(x,z)$ we get
$$
R_{1313}=(\frac{\partial^3 l}{\partial x^3}-3l\frac{\partial l}{\partial x}+
\frac{\partial l}{\partial z})y^2+
(\frac{\partial^2 m}{\partial x\partial z}-
$$
$$
2m\frac{\partial^2 l}{\partial x^2}-
l\frac{\partial^2 m}{\partial x^2}-
3\frac{\partial m}{\partial x}\frac{\partial m}{\partial x}-
\frac{\partial^2 l}{\partial x^2})y-
$$
$$
m\frac{\partial m}{\partial z}+
2m^2\frac{\partial l}{\partial x}+
m\frac{\partial l}{\partial x}+ml\frac{\partial m}{\partial x}
-m\frac{\partial m}{\partial z}+
$$
$$
2m^2\frac{\partial l}{\partial x}+
m\frac{\partial l}{\partial x}+ml\frac{\partial m}{\partial x},
$$
and
$$
R_{1323}=(-\frac{\partial m}{\partial z}+
2m\frac{\partial l}{\partial x}+
l\frac{\partial m}{\partial x}+
\frac{\partial l}{\partial x})/y
$$

    From the condition $R_{ijkl}=0$ it follows that the function $l(x,z)$
is the solution of the KdV-equation
$$
\frac{\partial^3 l}{\partial x^3}-3l\frac{\partial l}{\partial x}+
\frac{\partial l}{\partial z}=0.
$$
and all flat metrics of such type with help of solutions of  this
equation are determined.

   Note that after the Riemannian extensions of
the space with a given  metrics the metrics of the six-dimensional
space can be written. The equations of geodesics of such type of
6-dim space contains the linear second order ODE (Schrodinger
operator) which can be applied for integration of the KdV equation
and which is well known in theory of the KdV-equation.

\section{The applications for the Relativity}

   The notice of the Riemann extensions of a given metrics can be used for
the studying of general properties of the Riemannian spaces with the
Einstein conditions
$$
R_{ij}=g^{kl}R_{ijkl}=0
$$
on curvature tensor $R_{ijkl}$ and their generalizations.

     Let us consider some examples.

Let
$$
ds^2=-t^{2p_1}dx^2-t^{2p_2}dy^2-t^{2p_3}dz^2+dt^2   \eqno(14)
$$
be the metric of the Kasner type which has applications in
classical theory of gravitation.

   The Ricci tensor of this metrics has the components
$$
R_{ij}=\left(\begin{array}{llll}
\frac{(p_2+p_3+p_1-1)}{t^{2p_1-2}}&0&0&0\\
0&\frac{(p_2+p_3+p_1-1)}{t^{2p_2-2}}&0&0\\
0&0&\frac{p_3(p_2+p_3+p_1-1)}{t^{2p_3-2}}&0\\
0&0&0&\frac{(p_2+p_3+p_1-p_1^2-p_2^2-p_3^2)}{t^2}
\end{array}\right),
$$
and in case $R_{ij}=0$ we get well known the Kasner solution of
the vacuum Einsten equations.

Now we shall apply the construction of Riemann extension for the
metrics (14). In result we get the eight-dimensional space with
local coordinates $(x,y,z,t,P,Q,R,S)$ and the metrics
$$
ds^2=-2\Gamma^k_{ij}\xi_k dx^idx^j+2dxdP
+2dydQ+2dzdR+2dtdS \eqno(15)
$$
were $\Gamma^k_{ij}$ are the Christoffel coefficients of the
metrics (14) and $\xi_k=(P,Q,R,S)$.

They are:
$$
\Gamma^4_{11}=p_1t^{2p_1-1},\quad\Gamma^4_{22}=p_2t^{2p_2-1},\quad
\Gamma^4_{33}=p_3t^{2p_3-1},
$$
$$
\Gamma^1_{14}=p_1/t\quad
\Gamma^2_{24}=p_2/t,\quad\Gamma^3_{34}=p_3/t
$$

As result we get the metrics of the space $K^8$ in form
$$
ds^2=-2p_1t^{2p_1-1}S dx^2-2p_2t^{2p_2-1}S dy^2-2p_3t^{2p_3-
1}S dz^2-
$$
$$
4p_1/tP dxdt-4p_2/tQ dydt-4p_3/tR dzdt+2dxdP+2dydQ+2dzdR+
2dt dS
$$

The  Ricci tensor $^{8}R_{ij}$  has the nonzero components
$$
R_{11}= 2p_1t^{2p_1-2}(p_1+p_2+p_3-1),\quad
R_{22}=2p_2t^{2p_2-2}(p_2+p_2+p_3-1),\quad
$$
$$
R_{33}=2p_3t^{2p_3-2}(p_2+p_2+p_3-1),\quad
R_{44}=2(p_2+p_2+p_3-p_1^2-p_2^2-p_3^2)/t^2
$$
which are the same with components of the Ricci tensor
$^{4}R_{ij}$ of the space $K^4$.

So the geometry of the Riemann space before and after
extension is the same.

   In turn the equations of geodesics of extended space
$$
\frac{d^2t}{ds^2}+p_1t^{2p_1-1}(\frac{dx}{ds})^2+p_2t^{2p_2-1}(\frac{dy}{ds})^2+
p_3t^{2p_3-1}(\frac{dz}{ds})^2=0,
$$
$$
\frac{d^2 x}{ds^2}+2\frac{p_1}{t}\frac{d
x}{ds}\frac{dt}{ds}=0,\quad \frac{d^2
y}{ds^2}+2\frac{p_2}{t}\frac{d y}{ds}\frac{dt}{ds}=0,\quad
\frac{d^2 z}{ds^2}+2\frac{p_3}{t}\frac{d z}{ds}\frac{dt}{ds}=0,
$$
$$
\frac{d^2R}{ds^2}-2\frac{p_3}{t}\frac{dt}{ds}\frac{dR}{ds}-
2p_3t^{2p_3-1}\frac{dz}{ds}\frac{dS}{ds} +
$$
$$
\left(2\frac{p_1p_3t^{2p_1-1}}{t}(\frac{dx}{ds})^2+
2\frac{p_2p_3t^{2p_2-1}}{t}(\frac{dy}{ds})^2)+
2\frac{p_3^2t^{2p_3-1}}{t}(\frac{dz}{ds})^2+
2\frac{p_3}{t^2}(\frac{dt}{ds})^2\right)R+2\frac{p_3t^{2p_3-1}}{t}
\frac{dz}{ds}\frac{dt}{ds}S =0,
$$
$$
\frac{d^2Q}{ds^2}-2\frac{p_2}{t}\frac{dt}{ds}\frac{dQ}{ds}-
2p_2t^{2p_2-1}\frac{dy}{ds}\frac{dS}{ds} +
$$
$$
\left(2\frac{p_1p_2t^{2p_1-1}}{t}(\frac{dx}{ds})^2+
2\frac{p_2p_3t^{2p_3-1}}{t}(\frac{dz}{ds})^2)+
2\frac{p_2^2t^{2p_2-1}}{t}(\frac{dy}{ds})^2+
2\frac{p_2}{t^2}(\frac{dt}{ds})^2\right)Q+2\frac{p_2t^{2p_2-1}}{t}
\frac{dy}{ds}\frac{dt}{ds}S =0,
$$
$$
\frac{d^2P}{ds^2}-2\frac{p_1}{t}\frac{dt}{ds}\frac{dP}{ds}-
2p_1t^{2p_1-1}\frac{dx}{ds}\frac{dS}{ds} +
$$
$$
\left(2\frac{p_1p_3t^{2p_1-1}}{t}(\frac{dz}{ds})^2+
2\frac{p_2p_1t^{2p_2-1}}{t}(\frac{dy}{ds})^2)+
2\frac{p_1^2t^{2p_1-1}}{t}(\frac{dx}{ds})^2+
2\frac{p_1}{t^2}(\frac{dt}{ds})^2\right)P+2\frac{p_1t^{2p_1-1}}{t}
\frac{dx}{ds}\frac{dt}{ds}S =0,
$$
$$
\frac{d^2S}{ds^2}-2\frac{p_3}{t}\frac{dz}{ds}\frac{dR}{ds}-
2\frac{p_2}{t}\frac{dy}{ds}\frac{dQ}{ds}-
2\frac{p_1}{t}\frac{dx}{ds}\frac{dP}{ds}+
\frac{4p_2^2}{t^2}\frac{dy}{ds}\frac{dt}{ds}Q+
\frac{4p_3^2}{t}\frac{dz}{ds}\frac{dt}{ds}R+
$$
$$
\left(\frac{p_1(2p_1-1)t^{2p_1-1}}{t}(\frac{dx}{ds})^2+
\frac{p_2(2p_2-1)t^{2p_2-1}}{t}(\frac{dy}{ds})^2+
\frac{p_3(2p_3-1)t^{2p_3-1}}{t}(\frac{dz}{ds})^2\right)S=0
$$
contain the linear  $4\times4$ matrix system of the second order
ODE's for the additional coordinates $(P,Q,R,S)$
$$
\frac{d^2 \Psi}{ds^2}=A(x,y,z,t)\frac{d\Psi}{ds}+B(x,y,z,t) \Psi.
$$
Here $A,B$ are the  $4\times4$ matrix-functions  depending on the
coordinates $(x,y,z,t)$. This fact allow us to use the methods of
soliton theory for the integration of the full system of geodesics
and the corresponding Einstein equations.

    Note that the signature of the space
$^{8}D$ is $0$, i.e. it has the form $(++++----)$. From this
follows that starting from the Riemann space with the Lorentz
signature $(---+)$ we get after the extension the additional
 subspace with local coordinates $P,Q,R,S$ having the signature $(-+++)$.

\begin{rem}

   For the Schwarzschild metrics
$$
g_{ij}=\left(\begin{array}{cccc} -\frac{1}{1-m/x}&0&0&0\\
0&-x^2&0&0\\
0&0&-x^2sin^2 y&0\\
0&0&0&1-\frac{m}{x}
\end{array}\right)
$$
\end{rem}
the Christoffell coefficients are
$$
\Gamma^1_{11}=\frac{m}{2x(x+m)},\quad\Gamma^1_{22}=-(x+m),
\Gamma^1_{33}=-(x+m)\sin^2
y,\quad\Gamma^1_{44}=-\frac{(x+m)m}{2x^3},
$$
$$
\Gamma^2_{12}=\frac{1}{x},\quad \Gamma^2_{33}=-\sin y \cos y,
\quad\Gamma^3_{13}=\frac{1}{x}, \quad\Gamma^3_{23}=\frac{\cos
y}{\sin y},\quad \Gamma^4_{14}=-\frac{m}{2x(x+m)}.
$$
the system (8) for the surfaces of translations
$x(u,v),y(u,v),z(u,v),t(u,v)$ is nonlinear.

    After the extension with the help of a new coordinates $(P,Q,R,S)$  we get
the $S^8$ space with the metrics
$$
ds^2=-2\Gamma^1_{11}Pdx^2-2\Gamma^1_{22}Pdy^2-2\Gamma^1_{33}Pdz^2-
2\Gamma^1_{44}Pdt^2-
$$
$$
2\Gamma^2_{33}Qdz^2-4\Gamma^2_{12}Qdxdy-4\Gamma^3_{13}Rdxdz-4\Gamma^3_{23}Rdy
dzdx-4\Gamma^4_{14}Sdxdt.
$$

    In this case the system (8) for the 4-dimensional submanifolds
    the linear subsystem of equations for the coordinates $(P,Q,R,S)$
is contained and so can be investigated.

\section{Anti-Self-Dual-Kahler metrics and the second order ODE's}

Here we discuss the relations of the equations (1) with theory
of the ASD-Kahler spaces [27].

   It is known that all ASD null Kahler metrics are locally given by
$$
 ds^2=-\Theta_{tt}dx^2+2\Theta_{zt}dxdy-\Theta_{zz}dy^2+dxdz+dydt
$$
where the function$\Theta(x,y,z,t)$ is the solution of the equation
$$
\Theta_{xz}+\Theta_{yt}+\Theta_{zz}\Theta_{tt}-\Theta_{zt}^2=\Lambda(x,y,z,t),
$$
$$
\Lambda_{xz}+\Lambda_{yt}+\Theta_{tt}\Lambda_{zz}+\Theta_{zz}\Lambda_{tt}-
2\Theta_{zt}\Lambda_{tz}=0.
$$

   This system of equations has the solution in form
$$
\Theta=-\frac{1}{6}a_1(x,y)z^3+\frac{1}{2}a_1(x,y)z^2t-\frac{1}{2}a_3(x,y)zt^2+
\frac{1}{6}a_4(x,y)t^3
$$
and lead to the metrics
$$
ds^2=2(za_3-ta_4)dx^2+4(za_2-ta_3)dxdy+2(za_1-ta_2)dy^2+2dxdz+2dydt
$$
with geodesics determined by the equation
\begin{equation}
  y''+a_{1}(x,y)y'^3+3a_{2}(x,y)y'^2+3a_{3}(x,y)y'+a_{4}(x,y)=0  \label{Lio}
\end{equation}

   In this case the coefficients $a_i(x,y)$ are not arbitrary
but satisfy the conditions
$$
L_1=\frac{\partial}{\partial y}(a_{4y}+3a_2a_4)-\frac{\partial}{\partial x}
(2a_{3y}-a_{2x}+a_1 a_4)-3a_3(2a_{3y}-a_{2x})-a_4a_{1x}=0 ,
$$
$$
L_2=\frac{\partial}{\partial x}(a_{1x}-3a_1a_3)+\frac{\partial}{\partial y}
(a_{3y}-2a_{2x}+a_1a_4)-3a_2(a_{3y}-2a_{2x})+a_1a_{4y}=0 .
$$

  According with the Liouvlle theory this means that such type of equations
can be transformed to the  equation
$$
y''=0
$$
with the help of the points transformations.

     Note that the conditions $L_1=0,\quad L_2=0$ are connected with the
integrable nonlinear p.d.e.( as the equation (7') for example) and from this
 we can get a lot examples of ASD-spaces.

\section{Dual equations and the Einstein-Weyl geometry in theory of second
order ODE's}

 In the theory of the second order ODE's
$$
y''=f(x,y,y')
$$
 we have the following fundamental diagram:
\[
\begin{array}{ccccc}
&  & F(x,y,a,b)=0 &  &  \\
& \swarrow \nearrow &  & \searrow \nwarrow &  \\
y^{\prime \prime }=f(x,y,y^{\prime }) &  &  &  & b^{\prime \prime
}=g(a,b,b^{\prime }) \\
&  &  &  &  \\
\Updownarrow &  &  &  & \Updownarrow \\
&  &  &  &  \\
M^{3}(x,y,y^{\prime }) &  & \Longleftrightarrow &  & N^{3}(a,b,b^{\prime })%
\end{array}%
\]%
which show the relations between a given second order ODE
$y''=f(x,y,y')$ its general integral $F(x,y,a,b)=0$ and so called
dual equation $b''=g(a,b,b')$ which can be obtained from general
integral when variables $x$ and $y$ as the parameters are
considered.

  In particular for the equations of type (1) the dual equation
$$
b''=g(a,b,b')     \eqno(16)
$$
has the function $g(a,b,b')$ satisfying  the partial differential
equation
$$
g_{aacc}+2cg_{abcc}+2gg_{accc}+c^2g_{bbcc}+2cgg_{bccc}+
$$
$$
+g^2g_{cccc}+(g_a+cg_b)g_{ccc}-4g_{abc}-4cg_{bbc} -cg_{c}g_{bcc}-
$$
$$
-3gg_{bcc}-g_cg_{acc}+ 4g_cg_{bc}-3g_bg_{cc}+6g_{bb} =0.
\eqno(16')
$$
Koppish(1905), Kaiser (1914).

   According to the E.Cartan the exp\-res\-sions on curvature of the space of
linear elements ($x$, $y$, $y^{\prime }$) connected with equation
(1)
\[
\Omega _{2}^{1}=a[\omega ^{2}\wedge \omega _{1}^{2}]\,,\quad
\Omega _{1}^{0}=b[\omega ^{1}\wedge \omega ^{2}]\,,\quad \Omega
_{2}^{0}=h[\omega ^{1}\wedge \omega ^{2}]+k[\omega ^{2}\wedge
\omega _{1}^{2}]\>.
\]%
where:
\[
a=-\frac{1}{6}\frac{\partial ^{4}f}{\partial y^{\prime }{}^{4}}\,,\quad h=%
\frac{\partial b}{\partial y^{\prime }}\,,\quad k=-\frac{\partial \mu }{%
\partial y^{\prime }}-\frac{1}{6}\frac{\partial ^{2}f}{\partial
^{2}y^{\prime }}\frac{\partial ^{3}f}{\partial ^{3}y^{\prime }}\,,
\]%
and
\begin{eqnarray*}
6b &=&f_{xxy^{\prime }y^{\prime }}+2y^{\prime }f_{xyy^{\prime
}y^{\prime }}+2ff_{xy^{\prime }y^{\prime }y^{\prime }}+y^{\prime
}{}^{2}f_{yyy^{\prime
}y^{\prime }}+2y^{\prime }ff_{yy^{\prime }y^{\prime }y^{\prime }} \\
&+&f^{2}f_{y^{\prime }y^{\prime }y^{\prime }y^{\prime
}}+(f_{x}+y^{\prime }f_{y})f_{y^{\prime }y^{\prime }y^{\prime
}}-4f_{xyy^{\prime }}-4y^{\prime
}f_{yyy^{\prime }}-y^{\prime }f_{y^{\prime }}f_{yy^{\prime }y^{\prime }} \\
&-&3ff_{yy^{\prime }y^{\prime }}-f_{y^{\prime }}f_{xy^{\prime
}y^{\prime }}+4f_{y^{\prime }}f_{yy^{\prime }}-3f_{y}f_{y^{\prime
}y^{\prime }}+6f_{yy}\>.
\end{eqnarray*}%
two types of equations are evolved naturally : the first type from
the con\-di\-tion $a=0$ and second type from the condition $b=0$.

The first condition $a=0$ the equation in form (1) is determined
and the
second condition lead to the equations (16) where the function $%
g(a,b,b^{'})$ satisfies the above p.d.e..

  The E.Cartan has also shown that the Einstein-Weyl 3-folds  parameterize the
families of curves of equation (16) which is dual to the equation
(1).

    Some examples of solutions of equation (16) were obtained first in [2].

   As example for the function
\[
g=a^{-\gamma }A(ca^{\gamma -1})
\]%
we get the equation
\[
\lbrack A+(\gamma -1)\xi ]^{2}A^{IV}+3(\gamma -2)[A+(\gamma -1)\xi
]A^{III}+(2-\gamma )A^{I}A^{II}+(\gamma ^{2}-5\gamma +6)A^{II}=0.
\]

One solution of this equation is
\[
A=(2-\gamma)[\xi(1+\xi^2)+(1+\xi^2)^{3/2}]+(1-\gamma)\xi
\]

This solution corresponds to the equation
\[
b^{\prime \prime }=\frac{1}{a}[b^{\prime }(1+b^{\prime
}{}^{2})+(1+b^{\prime }{}^{2})^{3/2}]
\]%
with General Integral
\[
F(x,y,a,b)=(y+b)^{2}+a^{2}-2ax=0
\]%
The dual equation in this case has the form
\[
y^{\prime \prime }=-\frac{1}{2x}(y^{\prime }{}^{3}+y^{\prime })
\]
\begin{rem}

      For  more general classes of the form-invariant equations
the notice of dual equation is introduced by analogous way.

     For example for the form-invariantly equation of the type
$$
P_{n}(b') b''-P_{n+3}(b')=0,
$$
where $P_{n}(b')$ are the polinomial in $b'$ degree $n$  with
coefficients depending from the variables $a,b$ the dual equation
$$
b''=g(a,b,b')
$$
has right part $g(a,b,b')$ in form of equation
$$
\left|\begin{array}{cccc}\psi_{n+4} &\psi_{n+3} &...&\psi_{4} \\
\psi_{n+5} &\psi_{n+4}&...&\psi_{5} \\
 .& .&...&.\\
\psi_{2n+4} &\psi_{2n+3} &...&\psi_{n+4}\end{array}\right|=0
$$
where the functions $\psi_i$ are determined with help of the
relations
$$
4!\psi_4=-\frac{d^2}{da^2}g_{cc}+4\frac{d}{da}g_{bc}-g_c(4g_{bc}-\frac{d}{da}g_{cc})+
3g_b g_{cc}-6g_{bb},
$$
$$
i\psi_i=\frac{d}{da}\psi_{i-1}-(i-3)g_c\psi_{i-1}+(i-5)g_b\psi_{i-2},\quad
i > 4
$$

      As example for equation
$$
2yy''-y'^4-y'^2=0
$$
with solution
$$
x=a(t+\sin t)+b,\quad y=a(1-\cos t)
$$
we have a dual equation
$$
b''=-\frac{1}{a}\tan(b'/2).
$$

     According to above formulaes at $n=1$ we get the values
$$
4!\psi_4=\frac{3}{2a^3}\tan\frac{c}{2}(1+\tan^2\frac{c}{2})^3,
$$
$$
5!\psi_5=-\frac{15}{4a^4}\tan\frac{c}{2}(1+\tan^2\frac{c}{2})^4,
$$
$$
6!\psi_6=\frac{90}{8a^5}\tan\frac{c}{2}(1+\tan^2\frac{c}{2})^5,
$$
and the relation
$$
\left|\begin{array}{cc}\psi_{5} &\psi_{4} \\
\psi_{6} &\psi_{5}
 \end{array}\right|=0
$$
or
$$
\psi_{5}^2-\psi_4 \psi_6=0
$$
is satisfied.
\end{rem}

     Here we consider some properties  of the Einstein-Weyl spaces [15].

   A Weyl space is smooth  manifold equipped with
a conformal metric $g_{ij}(x),$ and a symmetric connection
$$
G^k_{ij}=\Gamma^k_{ij}-\frac{1}{2}(\omega_i \delta^k_j+ \omega_j
\delta^k_i -\omega_l g^{kl}g_{ij})
$$
with condition on covariant derivation of the metrics
$$
D_i g_{kj}=\omega_i g_{kj}
$$
where $\omega_i(x)$ are the components of the vector field.

  The Weyl connection $G^k_{ij}$ has a curvature tensor $W^i_{jkl}$ and the
Ricci tensor $W^i_{jil}$, which is not symmetrical $W^i_{jil}\neq
W^i_{lij}$ in general case.

A Weyl space satisfying the Einstein condition
$$
\frac{1}{2}(W_{jl}+W_{lj})=\lambda(x) g_{jl}(x),
$$
with some function $\lambda(x)$, is called  the Einstein-Weyl
space.

Let us consider some examples.

  The components of Weyl connection of 3-dim space:
$$
ds^2=dx^2+dy^2+dz^2
$$
are
$$
2G_1=\left |\begin{array}{ccc}
- \omega_{1} & - \omega_{2} & - \omega_{3} \\
\omega_{2} & - \omega_{1} & 0 \\
\omega_{3} & 0 & - \omega_{1}
\end{array} \right | ,
2G_2=\left | \begin{array}{ccc}
-\omega_{2} &  \omega_{1} & 0 \\
-\omega_{1} & -\omega_{2} & -\omega_{3}\\
0 & \omega_{3} & -\omega_{2}
\end{array} \right |,
2G_3=\left | \begin{array}{ccc}
- \omega_{3} & 0 & \omega_{1} \\
0 & - \omega_{3} & \omega_{2} \\
- \omega_{1} & - \omega_{2} & - \omega_{3}
\end{array} \right |.
$$

From the equations of the Einstein-Weyl spaces
$$
W_{[ij]}=\frac{W_{ij}+W_{ji}}{2}= \lambda g_{ij}
$$
we get the system of equations
$$
\omega_{3x}+ \omega_{1z}+ \omega_1 \omega_3=0,\quad \omega_{3y}+
\omega_{2z}+ \omega_2 \omega_3=0, \quad \omega_{2x}+ \omega_{1y}+
\omega_1 \omega_2=0,
$$
$$
2 \omega_{1x}+ \omega_{2y}+ \omega_{3z}- \frac{ \omega_{2}^2+
\omega_{3}^2}{2}=2 \lambda,\quad 2 \omega_{2y}+ \omega_{1x}+
\omega_{3z}- \frac{ \omega_{1}^2+ \omega_{3}^2}{2}=2 \lambda,
$$
$$
2 \omega_{3z}+ \omega_{2y}+ \omega_{1x}- \frac{ \omega_{1}^2+
\omega_{2}^2}{2}=2 \lambda.
$$

   Note that the first three equations  lead to the Chazy equation [16]
$$
{R}''' + 2 R {R}''-3{R'}^2=0
$$
for the function
$$
R=R(x+y+z) = \omega_1 + \omega_2 + \omega_3
$$
where $\omega_i= \omega_i (x+y+z)$ and in general case they are
generalization of classical Chazy equation.

     Einstein-Weyl geometry of the metric
$g_{ij}=diag(1,-e^U,-e^U)$ and vector $\omega_i=(2U_z,0,0)$ is
determined by the solutions of equation [17]
$$
U_{xx}+U_{yy}=(e^ U)_{zz}.
$$
This equation is equivalent to the equation
$$U_{\tau}=(e^{U/2})_{z}
$$
(after substitution  $U=U(x+y=\tau,z)$) having many-valued
solutions.

    The consideration of the E-W structure for the metrics
$$
ds^2=dy^2-4dxdt-4U(x,y,t)dt^2
$$
lead to the dispersionless KP equation [18]
$$
(U_t-UU_x)_x=U_{yy}.
$$

    2. The Einstein-Weyl geometry of the four-dimensional  Minkovskii
space

$$
ds^2=dx^2+dy^2+dz^2-dt^2
$$

     The components of the Weyl connection are
$$
2G_1=\left |\begin{array}{cccc}
- \omega_{1} & - \omega_{2} & - \omega_{3} & - \omega_4 \\
\omega_{2} & - \omega_{1} & 0 & 0 \\
\omega_{3} & 0 & - \omega_{1} & 0 \\
- \omega_{4} & 0 & 0 & - \omega_{1}
\end{array} \right |,\quad
2G_2=\left |\begin{array}{cccc}
-\omega_{2} &  \omega_{1} & 0 & 0 \\
-\omega_{1} & - \omega_{2} & - \omega_{3} & - \omega_4 \\
0 & \omega_{3} & -\omega_{2} & 0 \\
0 & - \omega_{4} & 0 & - \omega_{2}
\end{array} \right|,
$$
$$
2G_3=\left |\begin{array}{cccc}
- \omega_{3} & 0 & \omega_{1} & 0 \\
0 & - \omega_{3} & \omega_{2} & 0 \\
- \omega_{1} & - \omega_{2} & - \omega_{3} & - \omega_4 \\
0 & 0 & - \omega_{4} & - \omega_{3}
\end{array} \right |,\quad
2G_4=\left |\begin{array}{cccc}
- \omega_{4} & 0 & 0 & \omega_{1} \\
0 & - \omega_{4} & 0 & - \omega_{2} \\
0 & 0 - \omega_{4} & - \omega_{3} \\
- \omega_{1} & - \omega_{2} & - \omega_{3} & - \omega_4
\end{array} \right |.
$$

     The Einstein-Weyl condition
$$
W_{[ij]}=\frac{W_{ij}+W_{ji}}{2}= \lambda g_{ij}
$$
where
$$
W_{ij}=W^l_{ilj}
$$
and
$$
W^k_{ilj}=\frac{\partial G^k_{ij}}{\partial x^l}- \frac{\partial
G^k_{il}}{\partial x^j}+G^k_{in}G^n_{lj}-G^k_{jn}G^n_{il}
$$
 lead to the system of equations
$$
\omega_{3x}+ \omega_{1z}+ \omega_1 \omega_3=0,\quad \omega_{3y}+
\omega_{2z}+ \omega_2 \omega_3=0,
$$
$$
\omega_{2x}+ \omega_{1y}+ \omega_1 \omega_2=0,\quad \omega_{4x}+
\omega_{1t}+ \omega_1 \omega_4=0,
$$
$$
\omega_{4y}+ \omega_{2t}+ \omega_2 \omega_4=0,\quad \omega_{4z}+
\omega_{3t}+ \omega_3 \omega_4=0,
$$
$$
3 \omega_{1x}+ \omega_{2y}+ \omega_{3z}- \omega_{4t}+
\omega_{4}^2- \omega_{2}^2- \omega_{3}^2=2 \lambda,
$$
$$
3 \omega_{2y}+ \omega_{1x}+ \omega_{3z}- \omega_{4t}+
\omega_{4}^2- \omega_{1}^2- \omega_{3}^2=2 \lambda,
$$
$$
3 \omega_{3z}+ \omega_{2y}+ \omega_{1x}- \omega_{4t}
+\omega_{4}^3-
 \omega_{1}^2- \omega_{2}^2=2 \lambda.
$$
$$
3 \omega_{4t}- \omega_{2y}- \omega_{1x}- \omega_{3z}
+\omega_{3}^2+
 \omega_{1}^2+ \omega_{2}^2=2 \lambda.
$$

\section{On the solutions of dual equations}

    Equation (9) can be written in compact form
$$
\frac{d^2 g_{cc}}{da^2}-g_{c} \frac{dg_{cc}}{da}-4
\frac{dg_{bc}}{da}+ 4g_c g_{bc}-3g_b g_{cc}+6g_{bb}=0 \eqno(17)
$$
with help of the operator
$$
\frac{d}{da}=\frac{\partial}{\partial a}+ c
\frac{\partial}{\partial b}+ g \frac{\partial}{\partial c}.
$$

   It has many types of
the reductions and the simplest of them are
$$
 g=c^{\alpha}\omega[ac^{\alpha-1}],\quad g=c^{\alpha}\omega[bc^{\alpha-2}],
\quad g=c^{\alpha}\omega[ac^{\alpha-1},bc^{\alpha-2}], \quad
g=a^{-\alpha}\omega[ca^{\alpha-1}],
$$
$$
\quad g=b^{1-2\alpha}\omega[cb^{\alpha-1}], \quad
g=a^{-1}\omega(c-b/a), \quad g=a^{-3}\omega[b/a,b-ac],\quad
 g=a^{\beta/\alpha-2}\omega[b^{\alpha}/a^{\beta},
c^{\alpha}/a^{\beta-\alpha}].
$$

   To integrate a corresponding  equations let us consider some particular cases

    1. $g=g(a,c)$

From the condition (17) we get
$$
\frac{d^2 g_{cc}}{da^2}-g_{c} \frac{dg_{cc}}{da}=0 \eqno (18)
$$
where
$$
\frac{d}{da}=\frac{\partial}{\partial a}+ g
\frac{\partial}{\partial c}.
$$

    Putting into (18) the relation
$$
g_{ac}=-gg_{cc}+\chi(g_c)
$$
we get the equation for $\chi(\xi),\quad \xi=g_c$
$$
\chi(\chi''-1)+(\chi'-\xi)^2=0.
$$
    It has the solutions
$$
\chi=\frac{1}{2}\xi^2, \quad \chi=\frac{1}{3}\xi^2
$$

      So we get two reductions of the equation (17)
$$
g_{ac}+gg_{cc}-\frac{g_{c}^2}{2}=0
$$
and
$$
g_{ac}+gg_{cc}-\frac{g_{c}^2}{3}=0.
$$
\begin{rem}
  The first reduction of equation (17) is followed from its
presentation in form
$$
g_{ac}+gg_{cc}-\frac{1}{2}{g_c}^2+cg_{bc}-2g_b=h,
$$
$$
h_{ac}+gh_{cc}-g_c h_c+ch_{bc}-3h_b=0
$$
and was considered in [3].

   In particular case $h=0$ we get one equation
$$
g_{ac}+gg_{cc}-\frac{1}{2}g_{c}^2+cg_{bc}-2g_{b}=0
$$
which is the equation (17) for the  function $g=g(a,c)$. It can be
integrated with help of Legendre transformation (see [3]).
\end{rem}

   The solutions of the equations of type
$$
u_{xy}=uu_{xx}+\epsilon u_{x}^2
$$
were constructed in [19]. The  work of [20] showed that they can
be present in form
$$
u=B'(y)+\int[A(z)-\epsilon y]^{(1-\epsilon)/ \epsilon} dz,
$$
$$
x=-B(y)+\int[A(z)-\epsilon y]^{1/ \epsilon} dz.
$$

    To integrate above equations we apply the parametric representation
$$
g=A(a)+U(a,\tau), \quad c=B(a)+V(a,\tau).
$$
Using the formulaes
$$
g_c=\frac{g_{\tau}}{c_{\tau}}, \quad g_{a}=g_{a}+g_{\tau}\tau_{a}
$$
we get after the substitution in (17) the conditions
$$
A(a)=\frac{d B}{d a}
$$
and
$$
U_{a \tau}-\left(\frac{V_{a} U_{\tau}}{V_{\tau}}\right)_{\tau}+ U
\left(\frac{ U_{\tau}}{V_{\tau}} \right)_{\tau} - \frac{1}{2}
\frac{U_{\tau}^2}{V_{\tau}}=0.
$$

     So we get one equation for two functions $U(a,\tau)$ and $V(a,\tau)$.
Any solution of this equation the solution  of equation (17) is
determined.

     Let us consider the examples.
$$
A=B=0, \quad U=2\tau-\frac{a\tau^2}{2}, \quad V=a\tau-2\ln(\tau)
$$

   Using the representation
$$
U=\tau \omega_{\tau}-\omega,\quad V=\omega_{\tau}
$$
it is possible to obtain others solutions of this equation.

      Last time the problem of integration of the dual equation
      with the
right part $g=g(a,b')$ as function of two variables $a$ and $b'$
was solved in work [28].

     Here we present the construction of the solutions of this type.

\begin{pr}
In case $h\neq 0$ and $g=g(a,c)$ the equation (17) is equivalent
the equation
$$
\Theta_a(\frac{\Theta_a}{\Theta_c})_{ccc}- \Theta_c(\frac{\Theta_a}{\Theta_c}%
)_{acc}=1        \eqno(19)
$$
where
\[
g=-\frac{\Theta_a}{\Theta_c}\quad h_c=\frac{1}{\Theta_c}
\]
\end{pr}

To integrate this equation we use the presentation
\[
c=\Omega(\Theta,a)
\]

     From the relations
\[
1=\Omega_{\Theta}\Theta_c, \quad
0=\Omega_{\Theta}\Theta_a+\Omega_c
\]
we get
\[
\Theta_c=\frac{1}{\Omega_{\Theta}},\quad \Theta_a=-\frac{\Omega_a}{%
\Omega_{\Theta}}
\]
and
\[
\frac{\Omega_a}{\Omega_{\Theta}}(\Omega_a)_{ccc}+ \frac{1}{\Omega_{\Theta}}%
(\Omega_a)_{cca}=1
\]

Now we get
\[
\Omega_{ac}=\frac{\Omega_{a \Theta}}{\Omega_{\Theta}}= (\ln
\Omega_{\Theta})_a=K,\quad \Omega_{acc}=\frac{K_{\Theta}}{\Omega_{\Theta}}%
,\quad
\]
\[
\Omega_{accc}=(\frac{K_{\Theta}}{\Omega_{\Theta}})_{\Theta}\frac{1}{%
\Omega_{\Theta}},\quad (\Omega_{acc})_a= (\frac{K_{\Theta}}{\Omega_{\Theta}}%
)_a-\frac{\Omega_a}{\Omega_{\Theta}} (\frac{K_{\Theta}}{\Omega_{\Theta}}%
)_{\Theta}
\]

As result the equation (19) takes the form
$$
\left[\frac{(\ln\Omega_{\Theta})_{a\Theta}}{\Omega_{\Theta}}\right]%
_a=\Omega_{\Theta}
$$
and can be integrated with under the substitution
\[
\Omega(\Theta,a)=\Lambda_a
\]

So we get the Abel type equation for the function
$\Lambda_{\Theta}$
$$
\Lambda_{\Theta\Theta}=\frac{1}{6}\Lambda_{\Theta}^3+
\alpha(\Theta)\Lambda_{\Theta}^2+
\beta(\Theta)\Lambda(\Theta)+\gamma(\Theta) \eqno(20)
$$
with arbitrary coefficients $\alpha, \beta, \gamma$.

Let us consider the examples.

1. $\alpha=\beta=\gamma=0$

The solution of equation (20) is
\[
\Lambda=A(a)-6\sqrt{B(a)-\frac{1}{3}\Theta}
\]
and we get
\[
c=A^{\prime}-\frac{3B^{\prime}}{\sqrt{B-\frac{1}{3}\Theta}}
\]
or
\[
\Theta=3B-27\frac{B'^2}{(c-A')^2}
\]
This solution corresponds to the equation
\[
b''=-\frac{\Theta_a}{\Theta_c}=-\frac{1}{18B'}
b'^3+\frac{A'}{6B'}b'^2+ (\frac{B''}{B'}-\frac{A'^2}{6B'}
)b'+A''+\frac{A'^3}{18B'}-\frac{A'B''}{B'}
\]
cubical on the first derivatives $b'$ with arbitrary coefficients $%
A(a),B(a)$. This equation is equivalent to the equation
\[
b''=0
\]
under the point transformation.

     In fact from the formulaes
$$
L_1=\frac{\partial}{\partial
y}(a_{4y}+3a_2a_4)-\frac{\partial}{\partial x} (2a_{3y}-a_{2x}+a_1
a_4)-3a_3(2a_{3y}-a_{2x})-a_4a_{1x} ,
$$
$$
L_2=\frac{\partial}{\partial
x}(a_{1x}-3a_1a_3)+\frac{\partial}{\partial y}
(a_{3y}-2a_{2x}+a_1a_4)-3a_2(a_{3y}-2a_{2x})+a_1a_{4y} .
$$
which are determined the components of a projective curvature of
the space of linear elements for the equations in form
$$
y''+a_{1}(x,y){y'}^3+3a_{2}(x,y){y'}^2+3a_{3}(x,y)y'+a_{4}(x,y)=0
$$
we have
$$
a_1(x,y)=\frac{1}{18 B'},\quad a_2(x,y)=-\frac{A'}{18 B'},\quad
a_3(x,y)=\frac{A'^2}{18 B'}-\frac{B''}{3 B'},\quad
a_4(x,y)=\frac{A'B''}{B'}-\frac{A'^3}{18 B'}-A''
$$
and conditions
$$
L_1=0,\quad L_2=0.
$$
are valid.

     This means that our equation determins a projective flat
structure in space of elements  $(x,y,y')$.
\begin{rem}

   The conditions
$$
L_1=0,\quad L_2=0.
$$
 correspond to the solutions of the equation (3) in form
$$
g(a,b,b')=A(a,b)b'^3+3B(a,b)b'^2+3C(a,b)b'+D(a,b).
$$
\end{rem}

\section
{The third-order ODE's and the the Weyl-geometry}

    In the works of E.Cartan the geometry of the equation
$$
b'''=g(a,b,b',b'')
$$
with General Integral in form
$$
F(a,b,X,Y,Z)=0
$$
has been studied.

   It has been shown  that there are a lot of types of geometrical structures
connected with this type of equations.

     Recently [21,22] the geometry of Third-order ODE's has been considered
in context of the null-surface formalism and it has been
discovered that in the case
 the function $g(a,b,b',b'')$ is satisfied the conditions:
$$
\frac{d^2 g_r}{da^2}-2g_{r} \frac{dg_r}{da}-3 \frac{dg_c}{da}+
\frac{4}{9}g_{r}^3 +2g_c g_r+6g_b=0 \eqno(21)
$$
$$
\frac{d^2 g_{rr}}{da^2}-\frac{d g_{cr}}{da}+g_{br}=0 \eqno(22)
$$
where
$$
\frac{d}{da}=\frac{\partial}{\partial a}+ c
\frac{\partial}{\partial b}+ r \frac{\partial}{\partial c}+ g
\frac{\partial}{\partial r}.
$$
the Einstein-Weyl geometry in space of initial values has been
realized.

    We present here some  solutions of the equations (21,22) which
are connected with theory of the second order ODE's.

   In the notations of E.Cartan we study the Third-order differential
equations
$$
y'''=F(x,y,y',y'')
$$
where the function $F$ is satisfied to the system of conditions
$$
\frac{d^2 F_2}{dx^2}-2F_{2} \frac{dF_2}{dx}-3 \frac{dF_1}{dx}+
\frac{4}{9}F_{2}^3 +2F_1 F_2+6F_0=0
$$
$$
\frac{d^2 F_{22}}{dx^2}-\frac{d F_{12}}{dx}+F_{02}=0
$$
where
$$
\frac{d}{dx}=\frac{\partial}{\partial x}+ y'
\frac{\partial}{\partial y}+ y'' \frac{\partial}{\partial y'}+ F
\frac{\partial}{\partial y''}.
$$

    In particular the third order equation
$$
y'''=\frac{3y'y''^2}{(1+y'^2)}
$$
of all cycles on the plane is a good example connected with the
Einstein-Weyl geometry.

   We consider the case of equations
$$
y'''=F(x,y',y'')
$$

In this case $F_0=0$ and from the second equation we have
$$
H_{x2}+y''H_{12}+FH_{22}=0
$$
where
$$
F_{x2}+FF_{22}-\frac{F_{2}^2}{2}+y''F_{12}-2F_1=H
$$

 With the help of this relation the first equation gives us the condition
$$
H_x+y''(H_1-F_{11})-FF_{12}-\frac{1}{18}F_2^3-F_{x1}=0
$$

In the case
$$
H=H(F_2), \quad and \quad F=F(x,y'')
$$
we get the condition on the function $F$
$$
F_{x2}+FF_{22}-\frac{F_2^2}{3}=0
$$

    The corresponding third-order equation is
$$
y'''=F(x,y'')
$$

and it is connected with the second-order  equation
$$
z''=g(x,z').
$$

       Another example is the solution of the system for the function
$F=F(x, y',y'')$ obeying to the equation
$$
F_{x2}+FF_{22}-\frac{F_{2}^2}{2}+y''F_{12}-2F_1=0
$$

    In this case $H=0$ and we get the system of equations
$$
F_{x2}+FF_{22}-\frac{F_{2}^2}{2}+y''F_{12}-2F_1=0
$$
$$
y''F_{11}+FF_{12}+\frac{1}{18}F_2^3+F_{x1}=0
$$
with the condition of compatibility
$$
(\frac{F_2^2}{6}-F_1)F_{22}+2F_2F_{12}+3F_{11}=0
$$

\section{Acknowledgement}

The author  thanks  the Cariplo Foundation (Center Landau-Volta,
Como, Italy), INTAS-99-01782 Programm, NATO-Grant, The Royal
Swedish Academy of Sciences for financial support.

{}


\begin{thebibliography}{}

\bibitem{1} V. Dryuma, {\it  Application of the E. Cartan method for
studyuing of nonlinear dynamical systems}, In ''Matematicheskie
issledovaniya, Kishinev, Stiintsa, 1987, v.92, 49-68,.
\bibitem{2} V. Dryuma, {\it Projective duality in theory of the second
order differential equations}, Mathematical Researches, Kishinev,
Stiintsa, 1990, v.112, 93--103.
\bibitem{3} Dryuma V.S., {\it On Initial values problem in theory of the
second order ODE's}, Proceedings of the  Workshop on Nonlineariyu,
Integrability and all that: Twenty years after NEEDS'79,
 Gallipoli(Lecce), Italy,  July 1-July 10, 1999, ed.  M.Boiti, L.
Martina, F. Pempinelli, B.Prinari and G.Soliani, World Scientific,
Singapore, 2000, 109-116.
\bibitem{4} V. Dryuma, {\it On Geometry of the second order differential
equations}, Proceedings of Conference Nonlinear Phenomena, ed.
K.V. Frolov,
 Moskow, Nauka, 1991, 41-48.
\bibitem{5} V. Dryuma, {\it Geometrical properties of multidimensional
differential equations and the Finsler metrics of dynamical
systems}, Theoretical and Mathematical Physics, Moskow, Nauka,
1994, v.99, no.2, 241-249.
\bibitem{6} Dryuma V.S., {\it Geometrical properties of nonlinear dynamical
systems}, Proceedings of the First Workshop on Nonlinear Physics,
Le Sirenuse,
 Gallipoli(Lecce), Italy June 29-July 7, 1995, ed. E. Alfinito, M.Boiti, L.
Martina and  F. Pempinelli, World Scientific, Singapore, 1996,
83-93.
\bibitem{7} R. Liouville, {\it Sur les invariants de
certaines \'{e}quations diff\'{e}rentielles et sur leurs
applications},
 J. de L'\'{E}cole Polytechnique {\bf 59}, 7-76 (1889).
\bibitem{8} A. Tresse, {\it D\'{e}termination des Invariants
ponctuels de l'\'{E}quation differentielle ordinaire de second
ordre: $y''=w(x,y,y') $}. Preisschriften der f\"urstlichen
Jablonowski'schen Gesellschaft XXXII, Leipzig, S. Hirzel, 1896.
\bibitem{9} A. Tresse, {\it Sur les invariants
diff\'{e}rentiels des groupes continus de transformations},
 Acta Math. {\bf 18}, 1-88 (1894).
\bibitem{10} E. Cartan, {\it Sur les vari\'{e}t\'{e}s a
connexion projective}, Bulletin de la Soci\'{e}t\'{e}
Math\'{e}mat. de France {\bf 52}, 205-241 (1924).
\bibitem{11} G.Thomsen, {\it \"{U}ber die topologischen
Invarianten der Differentialgleichung $
y''=f(x,y)y'^3+g(x,y)y'^2+h(x,y) y' +k(x,y) $}, Abhandlungen aus
dem mathematischen Seminar der Hamburgischen Universit\"at, {\bf
7}, 301-328, (1930).
\bibitem{12} Paterson E.M., Walker A.G.,{\it Riemann extensions}
Quart. J. Math. Oxford {\bf 3}, 19-28 (1952).
\bibitem{13} Wilczynski E.{\it Projective Differential Geometry of
Curved Surfaces}, Transaction of American Mathematical Society,
{\bf 9}, 103-128, (1908).
\bibitem {14} Cartan E.{\it Sur une classe d'espaces de Weyl}
 Ann. Ec. Norm. Sup. {\bf 14}, 1--16, (1943).
\bibitem {15} Pedersen H., Tod K.P.
{Three dimensional Einsten-Weyl Geometry}
 Advances in Mathematics {\bf 97}, 71--109, {1993}.
\bibitem{16} Chazy J. {\it Sur les equations differentielle don'tintegrale
possede un coupure essentielle mobile}, C.R. Acad. sc. Paris, {bf
150}, 456-458, (1910).
\bibitem{17} Ward R.{\it Einstein-Weyl spaces and Toda fields},
Classical and Quantum Gravitation, {\bf 7}, L45-L48, (1980).
\bibitem{18} Dunaisjski M., Mason L, Tod K.P.{\it Einstein-Weyl geometry,
the dKP equation and twistor theory}, arXiv:math. DG/0004031, 6
Apr.2000.
\bibitem{19} Calogero F.{\it A solvable nonlinear wave equation},
Studies in Applied mathematics, {\bf LXX}, N3, 189--199, (1984).
\bibitem{20} Pavlov M.{\it The Calogero equation and Liouville type
equations}, arXiv:nlin. SI/0101034, 19 Jan. 2001.
\bibitem{21} Tanimoto M.{\it On the null surface formalism},
arXiv:gr-qc/9703003, 1997.
\bibitem{22}D.M.Forni, M.Iriondo, C.N.Kozameh {\it Null surface formalism
in 3D}, arXiv:gr-qc/0005120, 26 May 2000.
\bibitem{23} Tod K.P.{\it Einstein-Weyl spaces and Third-order Differential
 equations}, J.Math.Phys, N9, 2000.
\bibitem{24} S.Frittelli, C.N.Kozameh, E.T.Newman {\it Differential geometry
from differential equations}, arXiv:gr-qc/0012058, 15 Dec. 2000.
\bibitem{25} Sprot J.C.{\it Symplest dissipative chaotic flow},
Physics Letters, A228 (1997), 271-274.
\bibitem{26} Konopelchenko B.{\it The non-abelian (1+1)-dimensional Toda
lattice}, Physics Letters, A156 (1991), 221-222.
\bibitem{27} Dunajski M.{\it Anti-self-dual four-manifolds with a
parallel spinors}, arXiv:math. DG/0102225, 18 Oct. 2001.1-18.
\bibitem{28} Dryuma V.,Pavlov M.{\it On initial value
problem in theory of the second order ODE}, Buletinul AS RM,
matematica,v.2(42), 2003, 51-58.
\bibitem{29} Dryuma V.{\it Applications of the Riemann and the Einstein-Weyl
geometry in theory of the second order ODE}, Theoretical and
Mathematical Physics , v.128, N1, 2001, 15-26.
\bibitem{30} Hietarinta J., Dryuma V.{\it Is my ODE a Painleve equation
in Disguise?},Journal of Nonlinear Mathematical Physics, v.9,
Suppl.1, 2002, 67-74.
\bibitem{31} Dryuma V. {\it Integration of the Einstein equations with tools
of linear problem in eight-dimensional space},Abstract, ICM,
August, 20-28, China, Bejijng, 2002, p.61.
\bibitem{32} Dryuma V.{\it On geometry of second order ODE's}, Report on
 the 16-NEEDS, Cadiz, Spain, June 10-15, 2002.
\end{thebibliography}
\end{document}